\documentclass{article}[12pt,a4paper]

\usepackage[pages=all, color=black, position={current page.south}, placement=bottom, scale=1, opacity=1, vshift=5mm]{background}
\SetBgContents{
} 

\usepackage[margin=1in]{geometry} 
\usepackage{graphicx,subcaption,lipsum}
\usepackage{amsmath}
\usepackage{amsthm}
\usepackage{amssymb}
\usepackage[english]{babel}
\addto{\extrasenglish}{%

}
\usepackage{subcaption}

\usepackage{lipsum}
\usepackage{setspace}

\usepackage{booktabs}
\usepackage{adjustbox}

\usepackage[rightcaption]{sidecap}
\graphicspath{ {fig2/} }

\usepackage[utf8]{inputenc}
\usepackage{hyperref}
\hypersetup{
unicode,
colorlinks,
breaklinks,
urlcolor=blue, 
linkcolor=blue, 
pdfauthor={Author One, Author Two, Author Three},
pdftitle={A simple article template},
pdfsubject={A simple article template},
pdfkeywords={article, template, simple},
pdfproducer={LaTeX},
pdfcreator={pdflatex}
}

\usepackage{hyperref}
\usepackage[numbers]{natbib}

\usepackage[nottoc]{tocbibind}

\usepackage{graphicx}
\theoremstyle{plain}

\theoremstyle{definition}

\usepackage{graphicx, color}
\usepackage{algorithm, algpseudocode} 
\usepackage{mathrsfs} 

\title{Hierarchical Bayesian Modeling of Total Column Ozone: Unraveling Equatorial Variability over Ethiopia Using Satellite Data and Multisource Covariates}

\author{Yassin Tesfaw Abebe$^{1,3}$ \and Abdu Mohammed Seid$^1$ \and Lassi Roininen$^{2}$ \and U. Jaya Parakash Raju$^{1}$ \and Abebaw Bizuneh Alemu$^{1}$
}

\date{
$^1$Bahir Dar University, Ethiopia  \\ \texttt{abdum442@yahoo.com}\\%
$^2$LUT University \\ \texttt{Lassi.Roininen@lut.fi}\\
$^3$Mekdela Amba University, Ethiopia \\ \texttt{ytesfaw@yahoo.com}\\[2ex]%
}

\begin{document}
\maketitle

\begin{abstract}
Understanding the spatiotemporal dynamics of total column ozone (TCO) is critical for monitoring ultraviolet (UV) exposure and ozone trends, particularly in equatorial regions where variability remains underexplored. This study investigates monthly TCO over Ethiopia (2012--2022) using a Bayesian hierarchical model implemented via Integrated Nested Laplace Approximation (INLA). The model incorporates nine environmental covariates, capturing meteorological, stratospheric, and topographic influences alongside spatiotemporal random effects. Spatial dependence is modeled using the Stochastic Partial Differential Equation (SPDE) approach, while temporal autocorrelation is handled through an autoregressive structure.
The model shows strong predictive accuracy, with correlation coefficients of 0.94 (training) and 0.91 (validation), and RMSE values of 3.91 DU and 4.45 DU, respectively. Solar radiation, stratospheric temperature, and the Quasi-Biennial Oscillation are positively associated with TCO, whereas surface temperature, precipitation, humidity, water vapor, and altitude exhibit negative associations. Random effects highlight persistent regional clusters and seasonal peaks during summer.
These findings provide new insights into regional ozone behavior over complex equatorial terrains, contributing to the understanding of the equatorial ozone paradox. The approach demonstrates the utility of combining satellite observations with environmental data in data-scarce regions, supporting improved UV risk monitoring and climate-informed policy planning.

\vspace{.25in}
\end{abstract}


\section{Introduction}
Global warming and climate change are among the most pressing challenges facing life on Earth, largely driven by the excessive release of greenhouse gases such as carbon dioxide \citep{staehelin2001ozone}. In this context, ozone plays a critical role in the global climate system. Ozone, a colorless, odorless reactive gas consisting of three oxygen atoms, occurs naturally in the troposphere and stratosphere. Approximately 90\% of atmospheric ozone forms the ozone layer in the stratosphere, 10-50 kilometers above Earth's surface \citep{staehelin2001ozone}. Solar ultraviolet (UV) radiation is absorbed by this layer, protecting life on Earth. The remaining 10\% is in the troposphere, extending from the surface to 10 kilometers above the earth's surface, depending on latitude. Ozone is a greenhouse gas and air contaminant in this lower layer. UV light splits oxygen molecules, allowing oxygen atoms to combine and generate stratospheric ozone. In contrast, in the troposphere, nitrogen oxides and volatile organic molecules combine with sunlight to form ozone. Main sources are industrial pollutants, motor vehicle exhaust, and chemical solvents. Ozone in the stratosphere protects living things from UV light that can cause skin cancer, cataracts, and immune system suppression. It safeguards marine life and agriculture. The troposphere's ozone is damaging \citep{murray2020global}. The gas destroys lung tissue, affects respiratory function, and increases asthma and bronchitis risks. It damages vegetation, creates urban smog, and lowers agricultural yield. It can damage forests and ecosystems over time. Tropospheric ozone absorbs Earth's infrared radiation to trap heat as a greenhouse gas. After carbon dioxide and methane, it is the third most important greenhouse gas. Therefore, investigating ozone's sources, distribution, and effects is crucial to understanding its involvement in regional and worldwide environmental change.

Total ozone at a specific location on Earth's surface refers to the cumulative amount of ozone present in the atmospheric column above that point, extending from the ground to the top of the atmosphere. This measure, known as Total Column Ozone (TCO), is typically expressed in Dobson Units (DU), where 1 DU corresponds to a 0.01 mm thick layer of pure ozone at standard temperature and pressure \citep{essd1338852021}. TCO is measured using ground-based instruments and satellite observations. Variations in TCO are influenced by multiple factors, including atmospheric circulation, chemical composition, solar radiation, and seasonal dynamics. Understanding these variations is essential for evaluating ozone-related environmental and health impacts, as well as for monitoring changes associated with climate variability.

Total ozone at a specific location on Earth's surface refers to the cumulative ozone content in the atmosphere above that point, representing the total ozone concentration within that vertical column. This is done using ground-based stations and satellites, and the amount is measured in Dobson Units (DU), where 1 DU equals a layer of ozone that is 0.01 mm thick at room temperature and pressure \citep{essd1338852021}. Total Column Ozone (TCO) refers to the total amount of ozone within a vertical column of air extending from the Earth's surface up to the top of the atmosphere. Weather patterns and changes in the atmospheric chemicals are just two factors that have an impact on the concentration and thickness of TCO. According to \citet{fahey2018scientific}, an ozone hole is likely to form in polar regions where the concentration is less than 220 DU. Ozone holes are associated with UV radiation from the sun, which is one of the primary causes of skin cancer \citep{saraiya2004interventions}. Because TCO plays a significant role in the Earth's climate system, it is essential to investigate its spatiotemporal variations in order to comprehend and forecast future environmental changes. When it comes to understanding the state of the atmosphere and finding solutions to environmental problems like climate change and ozone depletion, ozone monitoring is vital, and total column ozone is a key indicator of this layer's importance.

Over the last few decades, there has been a significant increase in research on ozone dynamics, emphasizing the interaction of topographic, climatic, anthropogenic, and natural factors on ozone dynamics. Altitude plays a crucial role in influencing the exchange between the stratosphere and the troposphere, affecting ozone dynamics. Higher elevations often lower TCO by mixing with ozone-poor tropospheric air, even though they are close to the ozone layer \citep{acp743112007, kuttippurath2023trends}. \citet{shangguan2019variability} found that surface temperature affects photochemical ozone production in the troposphere and lower stratosphere. Humidity and total column water vapor (TCWV) make ozone loss worse through HO\textsubscript{x} chemistry and radiative cooling \citep{nade2020intra}. According to \cite{lu2019meteorology}, when it rains, convection lifts pollutants like short-lived halogens that destroy ozone. On the other hand, wind speed (V-component) helps move ozone around by advection and changing the patterns of circulation in the stratosphere \citep{jana2014effect}. Incoming solar radiation (ISR) is the main photochemical factor that affects the variability of tropical ozone, and the rate of O\textsubscript{2} photolysis is determined by the intensity of UV-B light \citep{Nassif_2020}. The temperature of the stratosphere affects how quickly ozone is lost. Higher temperatures in the stratosphere stabilize ozone by turning reactive halogens into reservoirs \citep{shangguan2019variability}. Large-scale oscillations, like the Quasi-Biennial Oscillation (QBO), affect the transport of ozone around the equator. Westerly phases stop upwelling and help ozone stay in place \citep{GABIS20142499}. The analysis of these covariates' effect within a spatiotemporal framework yields insights into the complex and regionally varying behavior of TCO, especially in both topographically and climatically diverse regions like Ethiopia.

Most of the earlier studies that looked at TCO in space and time used seasonal and yearly changes in TCO and simple averages to find monthly, yearly, and inter-annual TCO changes to identify the highest and lowest occurrences of TCO levels \citep{rafiq2017long}. \cite{chen2014investigating} investigated the spatiotemporal variability of TCO over the Yangtze River Delta in China. They use the coefficient of relative variation to measure how the TCO changes over space in longitudinal and latitudinal bands. They also use a sinusoidal function to model seasonality as an annual cycle. Several studies have employed statistical and computational models to analyze TCO variability across different regions. \cite{10.1063/5.0143718} examined the spatiotemporal variations of TCO in Ethiopia, identifying higher ozone concentrations in equatorial regions compared to lower latitudes. These findings align with broader global studies that emphasize the influence of latitude and atmospheric circulation on ozone distribution. \cite{acp-22-15729-2022} also looked into how El Ni\~no-Southern Oscillation (ENSO) events affect stratospheric ozone and found that weather events cause big changes every year. But, in the latter two studies, they didn't study how other climatic and topographic variables affect TCO distribution. Looking at changes in ozone levels at the regional level helps us learn more about how the environment changes, which lets us make predictions using key covariates.

Bayesian spatiotemporal models, which use a hierarchy of sub-models to model complex environmental phenomena, are promising in air quality, pollution, and climate science \citep{cameletti2013spatio, fioravanti2021spatio}.  This approach incorporates explanatory variables to explain large-scale variability. It accounts for residual dependency using a space-time process and Gaussian random field (GRF). Dealing with huge amounts of data is hard to do on a computer when using the Bayesian model with a GRF. This is especially true when using complex spatial dependence measures like the Mat\'ern covariance function \citep{porcu2012advances}. Researchers have proposed a number of strategies to lessen the computing load associated with fitting complex temporal, spatial, and spatiotemporal models. According to \cite{lindgren2011explicit}, the stochastic partial differential equation (SPDE) method can be used to show a continuous Mat\'ern field with a sparse precision matrix and a discretely indexed Gaussian Markov random field (GMRF). This approach has good computational properties. People can do direct numerical calculations on the marginal posterior distributions without using the time-consuming Markov chain Monte Carlo (MCMC) simulations. Instead, they can use the Integrated Nested Laplace Approximation (INLA) algorithm \citep{rue2009approximate}. The R-INLA package (\url{ https://www.r-inla.org/}) allows for the fast and easy implementation of GMRF using the SPDE technique in a Bayesian hierarchical framework.

Even with these developments, there are still few local studies that concentrate on Ethiopia. Most of the studies that have been done so far only look at data from larger regions. They don't look at how the microclimate and geography of Ethiopia affect ozone dynamics \citep{asefa2020ethiopian}. Ethiopia's complex topography, ranging from the high-altitude highlands to the lowland places, makes it an ideal region to examine these influences in detail. Ethiopia’s geographical diversity and climatic variability \citep{fazzini2015climate} have attracted considerable attention in climate and environmental studies. With elevations ranging from 125 m below sea level to 4,620 m above sea level, the country's topography has a significant impact on its climate. This variation creates distinct climatic zones, including arid lowlands, temperate highlands, and regions with bimodal rainfall patterns. These climatic and topographical features make Ethiopia a unique natural laboratory for studying ozone dynamics. Despite this, limited research has been conducted on the spatiotemporal distribution of TCO in Ethiopia, leaving a critical gap in understanding ozone variability in this region. 

However, to our best knowledge, no study focuses on the spatiotemporal variability of TCO using a hierarchical spatiotemporal model and investigates the potential influence factors across Ethiopia. This study addresses this gap by using high-resolution gridded data to model TCO distribution over Ethiopia, incorporating meteorological, climatological, and geographical predictors to capture the intricate spatiotemporal patterns. This study uses monthly mean TCO data from the years 2012 to 2022 along with Bayesian spatiotemporal models to look at how TCO changes in Ethiopia, with a focus on key predictors. First, we set up a number of spatiotemporal models, and we check how well they can fit and predict the future using certain criteria on both the training and validation sets. Moreover, we analyze and identify key predictors influencing TCO variability using the selected model.

\subsection{Main contributions}
We model the spatiotemporal dynamics of TCO and its associated drivers using a hierarchical Bayesian framework, with inference performed via the INLA-SPDE approach. TCO is formulated as a geostatistical process with both fixed and random effect components. The fixed effects quantify the influence of meteorological, stratospheric, and topographic variables on TCO variability across Ethiopia. The random effects capture latent spatiotemporal structure in TCO, offering insights into variability beyond the scope of the fixed covariates. The main contributions of this paper are
\begin{enumerate}
\item Quantifying the effects of climatic and topographic variables on TCO while accounting for spatiotemporal variation, thereby identifying key factors of TCO variability over Ethiopia.
\item Developing and validating a spatiotemporal predictive model for TCO, capable of capturing both observed covariate effects and latent residual structure.
\item Highlighting the potential for future extensions using joint modeling approaches to better analyze and interpret the complex interactions influencing TCO distribution.
\end{enumerate}
In the following sections, we outline the structure of the paper. We show the dataset for the primary explanatory and response variables in Section \ref{data}. In Section \ref{model}, we develop models that consider both space and time using a Bayesian approach for TCO. In Section \ref{result}, we present the key results, summarize the model briefly, and discuss the significant factors influencing TCO in Ethiopia along with the model's predictions. In Section \ref{conclusion}, we conclude the study by summarizing our findings and exploring potential future research directions. 


\section{Data}\label{data}
\subsection{Total column ozone and spatial domain}
Ethiopia is located in the northeastern part of Africa and lies approximately between $3^{\circ}$ N and $15^{\circ}$ N latitudes and $33^{\circ}$ E and $48^{\circ}$ E longitudes \citep{asefa2020ethiopian}. It is characterized by diverse topography and climate. Elevations range from 125 m below sea level in the Danakil Depression, one of the hottest and driest places on Earth, to 4,620 m above sea level at Ras Dashen, the highest peak in the country. The climate varies significantly with altitude, transitioning from arid lowlands to cold highlands. Ethiopia experiences four distinct seasons: summer (\textit{kiremt}) from June to August (JJA), characterized by heavy rainfall; spring (\textit{thedey}) from September to November (SON), a moderate period following the rains; winter (\textit{bega}) from December to February (DJF), marked by dry weather; and fall (\textit{belg}) from March to May (MAM), which serves as a pre-rainy period. Seasonal changes vary in timing across different regions of the country due to variations in latitude and topography.
\begin{figure}[h]
\centering
\includegraphics[width=.6\linewidth]{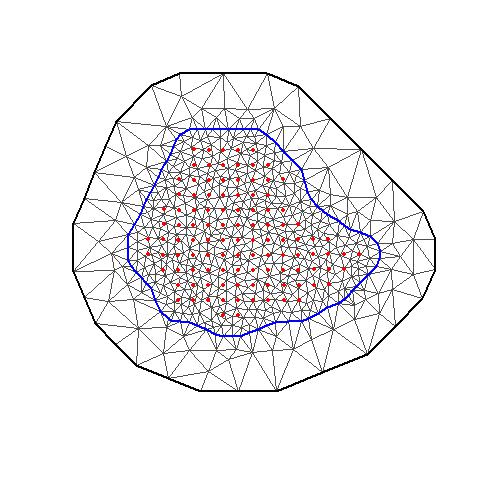}
\caption{The spatial domain for the study, together with the grid location (red circles), is used to construct the SPDE approximation to the continuous Mat\'ern field.}
\label{mesh}
\end{figure}
This complex interplay of elevation, latitude, and seasonality creates heterogeneous microclimates that significantly influence atmospheric chemistry. Consequently, Ethiopia’s variations make it an ideal region for studying spatiotemporal ozone dynamics. For this study, we used monthly Total Column Ozone (TCO) data for Ethiopia from the Ozone Mapping and Profiler Suite-Nadir Mapper (OMPS-NM, Version 2.1), developed by Ball Aerospace \& Technologies Corporation to measure ozone concentrations in the Earth’s atmosphere \citep{omps2002nm}. The data, accessed via the NASA Goddard Space Flight Center, cover an 11-year period from January 2012 to December 2022 with a spatial resolution of $1^{\circ}\times 1^{\circ}$. In our analysis, we looked at 108 specific locations in Ethiopia, which are shown as red circles on the mesh created for the SPDE approach in Figure \ref{mesh}.

\begin{figure}[H]
\centering
\includegraphics[width=.95\linewidth]{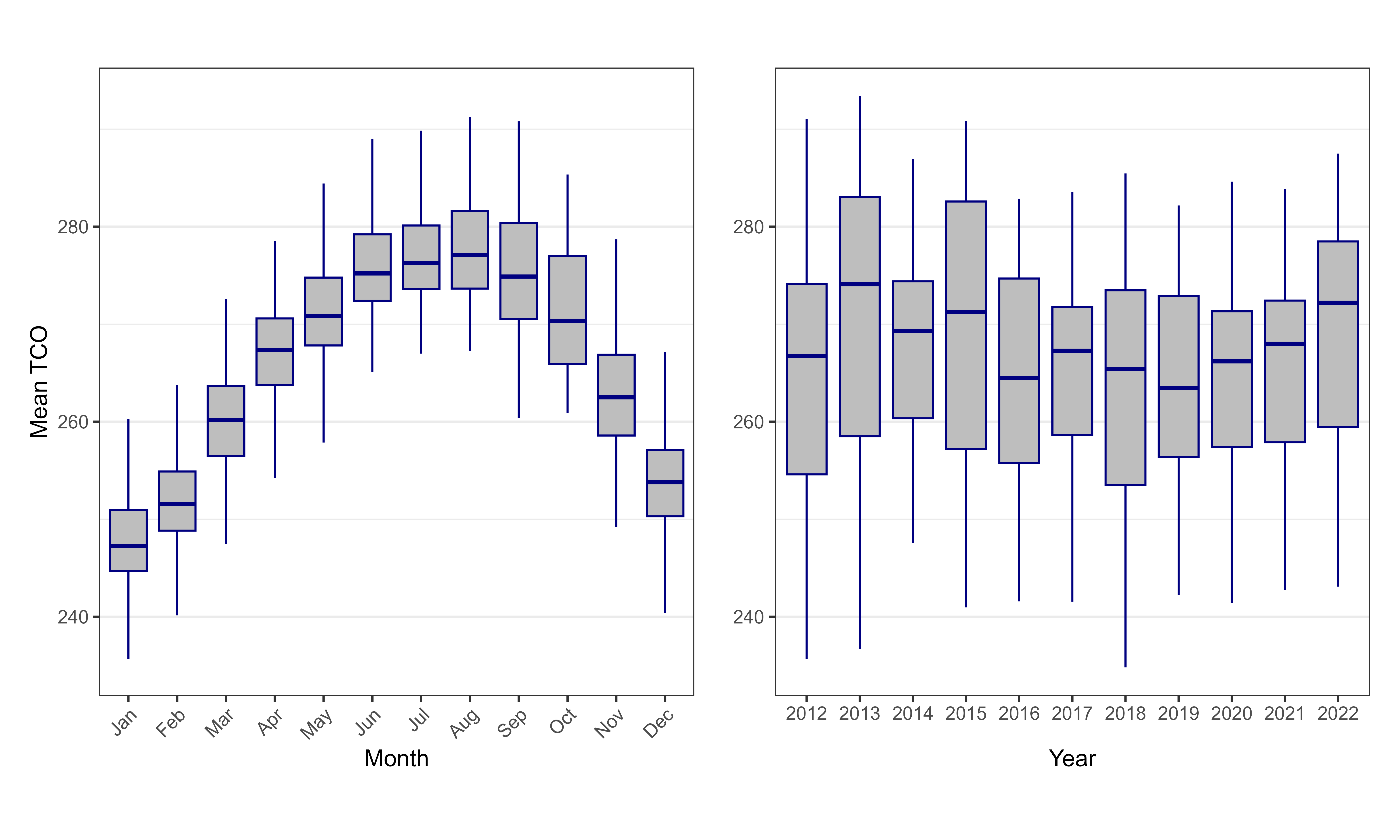}
\caption{Boxplots of averaged monthly variation TCO value (left) and averaged annual variation of TCO value (right) in Dobson unit (DU) over Ethiopia during 2012–2022}
\label{boxp}
\end{figure}

\begin{figure}[H]
\centering
\includegraphics[width=1\linewidth]{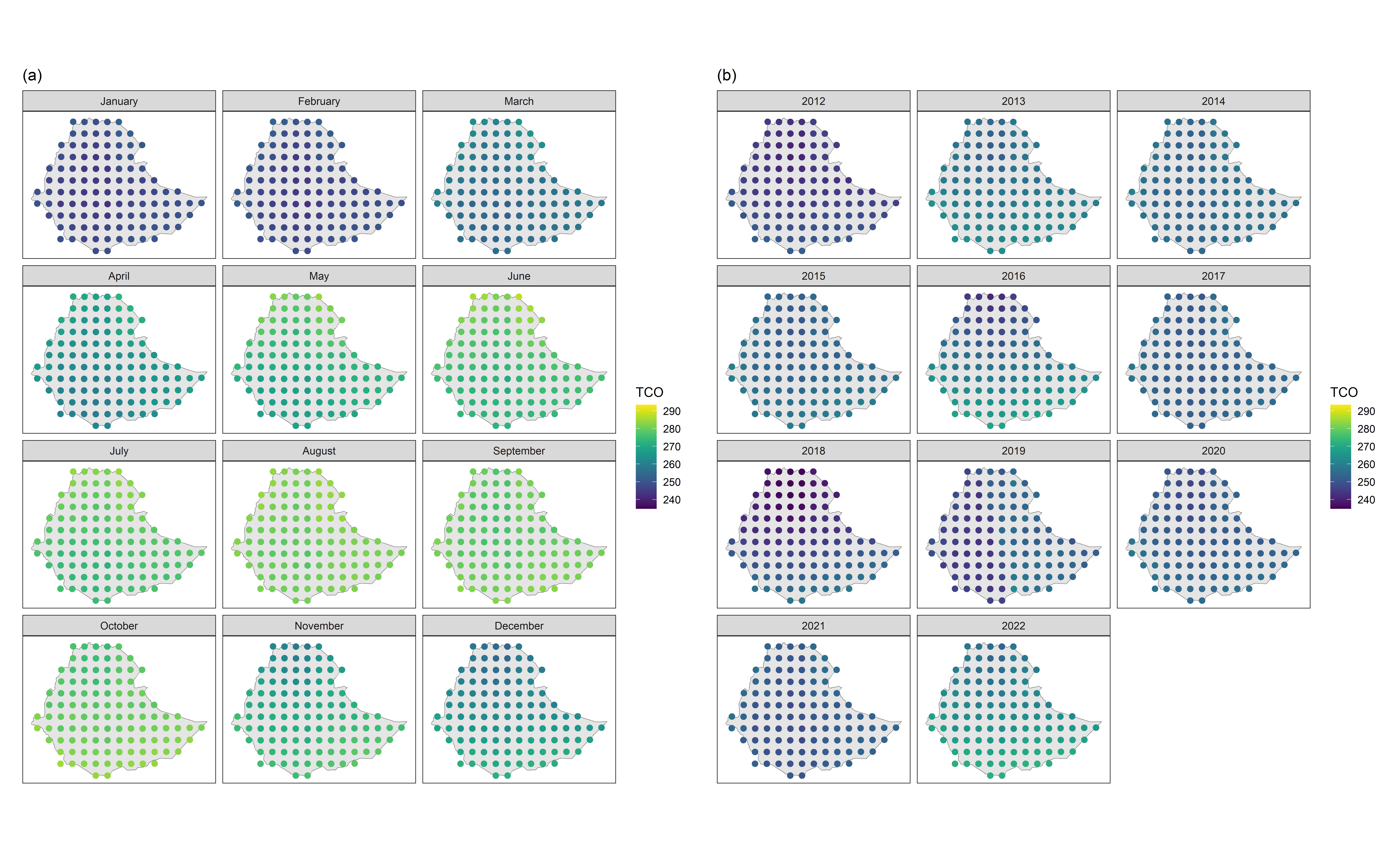} 
\caption{Spatial distribution of (a) monthly average TCO value and (b) yearly average TCO value at each grid location across Ethiopia.}
\label{spa_plot}
\end{figure}

We see that TCO has a noticeable peak that begins in late spring (May) and lasts through the summer and falls in winter. Figure \ref{boxp} (left) shows that TCO concentrations are comparatively lower in winter and higher during the summer. The mean spatial distribution of TCO in Ethiopia from 2012 to 2022 is displayed in Figure \ref{spa_plot}. Across all years, it is evident that the spatial mean TCO value across Ethiopia is almost similar, but in the northern part it is somehow relatively low, whereas the southern section has larger values that fall into different cluster regions. Because the TCO value changes in different places and times, we need to include this spatial and temporal dependence in our Bayesian hierarchical modeling.


\subsection{Explanatory variables}\label{exp_var}
Previous research on the temporal, spatial, and spatiotemporal variability of TCO led to the choice of a number of explanatory variables. We classify them into three groups: (1) meteorological variables, which show short-term changes in the atmosphere that affect TCO through photochemical and transport mechanisms; (2) climatological (stratospheric) variables, which show long-term or cyclical factors that cause ozone variability; and (3) geographical (topographic) variables, which show how ozone changes with elevation across Ethiopia's complex terrain. The rationale, data source, and relevance of each predictor are detailed below.

\subsubsection*{Meteorological variables} 
\textit{Surface temperature:} This study uses the monthly surface temperature \citep{hersbach2023era5} to show how temperature affects the photochemistry and vertical transport of ozone. In the context of Ethiopia’s diverse topography and climate zones, surface temperature exhibits notable spatial and seasonal variability, ranging from hot lowland regions (e.g., Afar and Somali regions) to cool highlands (e.g., central and northern) of Ethiopia. Elevated surface temperatures enhance convective activity, potentially lifting ozone-depleting pollutants into the upper troposphere, where they contribute to ozone loss \cite{ shangguan2019variability}. Additionally, surface warming can alter atmospheric stability and promote radiative cooling in the stratosphere, which suppresses ozone production \cite{NINGOMBAM2020104686}. This effect is particularly relevant in the lowland regions of Ethiopia, where persistent warming during the dry season and the period preceding the main rainy season may contribute to observed reductions in TCO.

\textit{Humidity:} Upper tropospheric lower stratospheric (UTLS) humidity, which is found in \cite{hersbach2023era5}, has a big effect on ozone chemistry by changing how clouds form and how NO\textsubscript{x} reacts with water \cite{nade2020intra}. In Ethiopia, humidity levels exhibit considerable seasonality, markedly rising during the primary rainy seasons. High humidity during this period can inhibit NO\textsubscript{x} induced ozone depletion by facilitating heterogeneous reactions on cloud particles. Simultaneously, increased humidity may augment the solubility and vertical transit of halogen compounds, complicating ozone fluctuation in Ethiopia's rainy regions.

\textit{Precipitation:} Monthly precipitation totals, from \cite{hersbach2023era5}, drive convective transport of ozone-depleting substances. Heavy rainfall during the June–September main rainy season enhances vertical injection of short-lived halogen compounds into the lower stratosphere. Additionally, the radiative shielding caused by heavy rainfall and the resulting cloud cover temporarily reduces photochemical ozone production \cite{jana2014effect}.

\textit{Total column water vapor:} Total column water vapor (TCWV) \cite{hersbach2023era5} modulates ozone through HO\textsubscript{x} mediated destruction and associated radiative effects. Seasonal TCWV peaks (July-September) over western Ethiopia coincide with the northward shift of the Intertropical Convergence Zone (ITCZ) \cite{nade2020intra}, which contributes to ozone stabilization via infrared cooling.

\textit{Wind speed:} The V-component wind at 30 hPa, derived from \cite{hersbach2023era5}, plays a key role in regulating cross-latitudinal ozone transport in the lower stratosphere. Over Ethiopia, this wind component is particularly influential during boreal winter (DJF), when stratospheric circulation patterns modulate the latitudinal distribution of ozone. Anomalous northward winds during this season can advect ozone-rich air masses from subtropical and mid-latitude regions into Ethiopia, increasing ozone concentrations. Conversely, enhanced southward flow facilitates the intrusion of ozone-poor tropical air, leading to localized ozone reductions. Spatial heterogeneity in wind patterns can thus drive regional differences in ozone variability across Ethiopia's northern and southern sectors.

\subsubsection*{Climatological variables}
\textit{Quasi-Biennial Oscillation:} The Quasi-Biennial Oscillation (QBO), obtained from (\url{https://www.cpc.ncep.noaa.gov/}), is the main type of tropical stratospheric variability. Every 24 to 32 months on average, it is identifiable by shifting easterly and westerly wind patterns. These wind regimes descend progressively through the stratosphere and drive opposing vertical motions in the tropics and extratropics, along with meridional transport across latitudes. Through these dynamic effects, the QBO influences both the vertical and horizontal distribution of ozone. Over Ethiopia, which lies in the deep tropics, QBO phases modulate ozone via large-scale circulation changes \cite{GABIS20142499}. During easterly phases, ozone-poor air from the upper stratosphere tends to fall to the ground, and upwelling speeds up, which lowers the amount of ozone in the air. In contrast, westerly phases suppress tropical upwelling and facilitate ozone accumulation in the lower stratosphere, resulting in higher ozone values \cite{wang2022zonally}.

\textit{Stratospheric temperature:} Monthly temperatures at 30 hPa \cite{hersbach2023era5} are included to account for the thermal modulation of catalytic ozone destruction. Cooling trends in the lower stratosphere have been linked to delayed ozone recovery, particularly over high-altitude regions \cite{fahey2018scientific}. Colder stratospheric conditions slow down ozone-depleting reactions involving halogen and nitrogen compounds, leading to lower TCO values. This inverse relationship has been observed in other high-elevation and tropical regions such as the Tibetan Plateau, where lower-stratospheric temperature anomalies correlate strongly with ozone concentrations \cite{li2020analysis}. Given Ethiopia's varied topography, including extensive highland areas, stratospheric temperature serves as a key covariate in modeling TCO variability.

\textit{Incoming Solar Radiation:} Incoming solar radiation (ISR), from \cite{hersbach2023era5}, makes ozone in the stratosphere by breaking up O\textsubscript{2} molecules and making it easier for reactions to happen that make ozone \cite{jana2014effect}. Over Ethiopia, changes in the sun's zenith angle with the seasons and the ITCZ moving north cause big changes in the amount of solar radiation reaching the surface and stratosphere every month. These changes affect TCO: when the sun is shining the brightest (like in April and May), more UV-B light helps make ozone, but when the sun isn't shining as brightly, less ISR can make ozone loss worse across equatorial latitudes. Incorporating ISR into the model captures both the seasonal and annual drivers of TCO variability over Ethiopia’s tropical domain.

\subsubsection*{Topographic variable}
\textit{Altitude:} Ethiopia's complex terrain, ranging from the Danakil Depression (125 meters below sea level) to the Semien Mountains (over 4500 meters above sea level), plays a significant role in modulating ozone dynamics. Higher altitudes are associated with stronger stratosphere-troposphere exchange, increasing the likelihood of reaching ozone-poor air. In addition, elevation influences UV intensity, which in turn affects ozone production and destruction rates. Elevation data are obtained from the Global Multi-resolution Terrain Elevation dataset \citep{danielson2011global}.

A complete summary of all predictor variables, including data sources and descriptive statistics, is provided in Table~\ref{tab_pred}.

\begin{table}[H]
\centering
\begin{adjustbox}{width=\textwidth}
\begin{tabular}{llccc}
\toprule
\textbf{Variable (units)} & \textbf{Description} & \textbf{Range} & \textbf{Mean (Std)}& \textbf{Spatial resolution} \\
\midrule
TCO (Dobson unit) & Monthly average total column ozone & 234.8-293.4 & 266.1 (11.5)& $1^{\circ}\times1^{\circ}$   \\
Temperature (Kelvin) & Monthly average daily maximum temperature of air & & & \\
& at 2 m above the surface of the land on daily basis & 287.2-313.8& 299.3 (4.71) &  $0.25^{\circ}\times 0.25^{\circ}$ \\
Precipitation (mm) & Monthly mean accumulated precipitation & 0-55.3 &3.05 (0.16) & $0.25^{\circ}\times 0.25^{\circ}$  \\
Total column water vapor ($kg/m^2$) & Monthly average daily total column vertically integrated &  & &\\
& water vapor on daily basis & 5.60-53.17 & 27.62 (8.83) & $0.25^{\circ}\times 0.25^{\circ}$  \\
Incoming solar radiation ($W/m^2$) & Monthly average incoming solar flux & 338.3-441.9  & 411.9 (22.09) & $1^{\circ}\times 1^{\circ}$  \\
Humidity (\%) & Monthly average relative humidity at pressure level & & & \\
&ranging from 300 to 200 hPa & 15.5-89.5 & 44.35 (14.75) &$1^{\circ}\times 1^{\circ}$\\
Stratospheric temperature (Kelvin) & Monthly average daily temperature of the atmosphere at &  &  &   \\
&pressure levels ranging from 100 to 30 hPa on daily basis & 206.8-213.8 & 210.9 (1.43) & $0.25^{\circ}\times 0.25^{\circ}$  \\
Wind speed (m/s) & Monthly average V-component of wind speed of air  & &  & \\
&at altitude of 30 hPa in the stratosphere & -5.11-2.27 &-0.83 (0.88) & $0.25^{\circ}\times 0.25^{\circ}$  \\
Quasi-biennial oscillation ($m/s$) & Monthly average zonal winds at 30 hPa  & -29.1-15.09 & -3.213 (14.77) & - \\
Altitude (m) & Altitude of gridded location used for observation & -55-2848 & 1163.3 (660.64) & $30km\times 30km$  \\
\bottomrule
\end{tabular}
\end{adjustbox}
\caption{Description and summary statistics of the response and the spatial (or spatiotemporal) predictor variables}
\label{tab_pred}
\end{table}

The continuous predictors in Table \ref{tab_pred} are very different in size and range, so we make sure that all of the variables have the same spatial resolution $1^{\circ}\times 1^{\circ}$ and use monthly average values, which are the same as TCO data. Furthermore, we standardized each predictor to a mean of 0 and a standard deviation of 1 to prevent numerical issues. The transformation does not change the correlation and fitting results. Another important criterion for a multivariate regression model is that covariates should not be highly correlated with each other, and we check this using correlation analysis \cite{liang1993regression} and the variance inflation factor (VIF) \cite{Salmerón13082018}. In our analysis, all covariates are included, and the correlation coefficient is less than 0.75, and the VIF is evaluated to ensure it remains below 5 \cite{akinwande2015variance}, which is a common threshold used to mitigate issues related to multicollinearity.
%


\section{Spatiotemporal models} \label{model}
In this scheme, data from different locations are assumed to be realizations of a continuously indexed spatiotemporal process or random field $Y(\cdot,\cdot)$ evolving over time. Let $y(s_i, t)$ denote the mean TCO value at spatial location $s_i \in D$, where $D \subset \mathbb{R}^2$ is the study region and $t = 1,\ldots,T$ indexes the time period (2012–2022). We establish the following measurement equation:
\begin{equation}\label{a}
	y(s_i, t) = \eta(s_i,t) + \epsilon(s_i, t)
\end{equation}
where $\epsilon(s_i, t) \sim \mathcal{N}(0, \sigma_\epsilon^2)$ represents Gaussian measurement error assumed temporally and spatially uncorrelated and accounts for fine-scale variability, known in geostatistical literature as the nugget effect \cite{cressie2015statistics}. The latent process $\eta(s_i,t)$ represents the underlying true TCO value and is modeled to flexibly capture spatiotemporal dependence structures. Inspired by hierarchical Bayesian spatiotemporal models in air quality, ozone, and trace gas emission studies (\citet{sahu2012hierarchical}, \citet{cameletti2013spatio}, \citet{nicolis2019bayesian}, \citet{western2020bayesian}, \citet{otto2024spatiotemporal}), we define the latent process as:
\begin{equation}\label{b}
	\eta(\mathbf{s},t) = \mathbf{z}(\mathbf{s},t)\boldsymbol{\beta} + \xi(\mathbf{s},t)
\end{equation}
where $\mathbf{z}(\mathbf{s},t) = (z_0(s,t), z_1(s,t), \ldots, z_p(s,t))^\top$ is a vector of covariates that may vary in space and/or time, including meteorological, stratospheric, and topographic variables (see Section \ref{exp_var}), and $\boldsymbol{\beta} \in \mathbb{R}^{p+1}$ is the corresponding vector of coefficients. The term $\xi(\mathbf{s},t)$ captures residual spatiotemporal variability not explained by the covariates. We model $\xi(\mathbf{s},t)$ as a first-order autoregressive (AR(1)) process in time as
\begin{equation}\label{e}
	\xi(s_i,t) = \phi\xi(s_i,t-1) + \omega(s_i,t),
\end{equation}
with $\rho \in (-1,1)$ controlling temporal dependence, and $\xi(s_i, 1) \sim \mathcal{N}(0, \sigma_\omega^2 / (1 - \rho^2))$. The innovations $\omega(s,t)$ are spatially correlated Gaussian processes with Mat\'ern covariance is given as
\begin{equation}\label{3}
\mathrm{Cov}(\omega(s_i, t),\omega(s_j, t')) = 
\begin{cases}
\frac{\sigma^2_{\omega}}{2^{\nu-1}\Gamma(\nu)}(\kappa h)^\nu K_\nu(\kappa h)  &  t = t'\\
0 &  t \neq t'
\end{cases}
\end{equation}
where $h = \Vert s_i - s_j\Vert$ is the Euclidean distance between locations, $K_\nu$ is the modified Bessel function of the second kind of order, $\nu > 0, \ \sigma_\omega^2$ is the marginal variance, and $\kappa > 0$ is a scaling parameter related to the spatial range $r$ by $r = \sqrt{8\nu} / \kappa$, the distance at which spatial correlation drops to near 0.1 \cite{lindgren2011explicit, lindgren2015bayesian}. 
The above specification in Equation \eqref{3}, when $t = t'$, the covariance has a Mat\'ern structure that depends on $h$ between locations. On the other hand, when $t\neq t'$ the innovation term $\omega(\cdot,\cdot)$ has zero covariance, i.e., it does not contribute to the spatial dependence between locations for measurements taken at different times. The covariance function assumes the Gaussian field is second-order stationary and isotropic between locations \cite{lindgren2011explicit}.

The latent process $\eta(\mathbf{s},t)$ thus incorporates fixed effects, temporal correlation, spatial correlation, and spatiotemporal interaction through $\xi(\mathbf{s},t)$. To evaluate the contribution of each component, we define three hierarchically nested models based on restrictions on $\xi(\mathbf{s}, t)$ as follows.

\subsubsection*{Model 1: Covariate-only model (no random effects)}
This model includes only the fixed effects. It assumes $\xi(\mathbf{s}, t) = 0$, reducing Equation \eqref{b} to:
\begin{equation}\label{c}
\begin{split}
y(\mathbf{s}_i, t) &= \eta(\mathbf{s}_i, t) + \epsilon(\mathbf{s}_i, t) \\
\eta(\mathbf{s}_i, t) &= \mathbf{z}(\mathbf{s}_i, t)^\top \boldsymbol{\beta}.
\end{split}
\end{equation}
This model provides a baseline point for assessing the role of spatial and temporal random components.

\subsubsection*{Model 2: Additive spatial and temporal random effects (no interaction)} 
Here, \(\xi(\mathbf{s},t)\) is decomposed into independent spatial and temporal components:
\begin{equation}\label{5}
\begin{split}
y(\mathbf{s}_i, t) &= \eta(\mathbf{s}_i, t) + \epsilon(\mathbf{s}_i, t) \\
\eta(\mathbf{s}_i, t) &= \mathbf{z}(\mathbf{s}_i, t)^\top \boldsymbol{\beta} + f(t) + \omega(\mathbf{s}_i),
\end{split}
\end{equation}
where $f(t)$ is a spatially constant AR(1) temporal effect, and  $\omega(\mathbf{s}_i) $ is a zero-mean spatial Gaussian process independent of time. This model captures marginal spatial and temporal variation, but no spatiotemporal interaction.

\subsubsection*{Model 3: Full spatiotemporal model (primary model)} 
This is the most general formulation, allowing for dynamic spatiotemporal interaction, with
\begin{equation}\label{1}
\begin{split}
y(\mathbf{s}_i, t) &= \mathbf{z}(\mathbf{s}_i, t)^\top \boldsymbol{\beta} + \xi(\mathbf{s}_i, t) + \epsilon(\mathbf{s}_i, t), \\
\xi(\mathbf{s}_i, t) &= \rho \xi(\mathbf{s}_i, t-1) + \omega(\mathbf{s}_i, t),
\end{split}
\end{equation}
with $\omega(\mathbf{s}_i, t)$ defined as in Equation \eqref{3}. This model allows temporal evolution of the spatial field, enabling the latent process to exhibit both persistence and changing spatial structure over time. 
To perform Bayesian inference on the proposed spatiotemporal hierarchical models, particularly those involving spatial and spatiotemporal random fields, we employ the Integrated Nested Laplace Approximation (INLA) framework alongside the Stochastic Partial Differential Equation (SPDE) approach. This INLA-SPDE methodology is especially well-suited for latent Gaussian models with structured spatial components, such as those defined in Model 2 and Model 3. It enables accurate and scalable inference for complex spatiotemporal processes through deterministic approximations and efficient numerical integration.


\subsection{Inference}
\citet{rue2009approximate} proposed a computationally efficient approach called INLA for estimating posterior distributions in a broad class of latent Gaussian models. Unlike Markov chain Monte Carlo (MCMC) methods, INLA employs deterministic approximation techniques, including Laplace approximations and numerical integration, which allow for fast approximate Bayesian inference. \citet{lindgren2011explicit} extended the capabilities of INLA by linking Gaussian random fields (GRFs) and Gaussian Markov random fields (GMRFs) via the SPDE approach, demonstrating its effectiveness for complex spatial processes.

The latent process $\omega(\mathbf{s}, t)$ in Equation \eqref{e}, with a Mat\'ern covariance function of the form in Equation \eqref{3}, is the solution to a specific SPDE. This SPDE can be discretized as a GMRF using a finite basis function expansion defined over a triangulation of the study domain. The SPDE that defines the spatial field is given by
\begin{equation}\label{spde_eq}
\left(\kappa^2 - \Delta\right)^{\frac{\alpha}{2}} \tau\omega(\mathbf{s},t) = W(\mathbf{s},t), \qquad \alpha = \nu + \frac{d}{2}, \quad \kappa, \nu > 0,
\end{equation}
where $d = 2$ for $\mathbf{s} \in D \subset \mathbb{R}^2$, and $\Delta$ is the Laplacian operator. The parameter $\kappa$ controls the spatial scale, while $\alpha$ determines the smoothness of the GRF. The marginal variance of the GRF, $\sigma_\omega^2$, is given by $\sigma_\omega^2 = \frac{1}{(4\pi)\kappa^2 \tau_\omega^2}$. The term $W(\mathbf{s}, t)$ denotes Gaussian spatial white noise with unit variance. 

The solution to the SPDE in Equation~\eqref{spde_eq} is approximated by a finite element method (FEM) using piecewise linear basis functions defined over the triangulation of $D$ by
\begin{equation}\label{fem}
\omega(\mathbf{s},t) = \sum_{i=1}^{m}\psi(\mathbf{s})\tilde{w}_{it},
\end{equation}
where $m$ is the number of vertices in the triangulation, $\{\psi_i(\cdot)\}_{i=1}^m$ are the FEM basis functions, and $\{\tilde{w}_{it}\}$ are zero-mean Gaussian weights. The basis functions are linear over each triangle: $\psi_i(\mathbf{s})$ are 1 at the vertex $i$ and 0 at all other vertices. Values at interior points are determined by linear interpolation. While the Gaussian weights $\tilde{w}_{it}$ evolve over time, the basis functions remain fixed due to the static mesh. For each time point $t = 1, \dots, T$, the spatial field $\boldsymbol{\tilde{\omega}}_t := \omega(\cdot,t)$ follows a GMRF distribution: 
\begin{equation*}
\boldsymbol{\tilde{\omega}}_t \sim \mathcal{N}(\boldsymbol{0}, \boldsymbol{Q}_s^{-1}),
\end{equation*}
where $\boldsymbol{Q}_s$ is the $m$-dimensional spatial precision matrix, which is constant across time due to the assumed temporal independence in Equation \eqref{3}. Letting $\boldsymbol{\xi}_t := \xi(\cdot,t)$, the spatiotemporal dynamics are modeled as:
\begin{equation}\label{ar}
\boldsymbol{\xi}_t = \rho \boldsymbol{\xi}_{t-1} + \boldsymbol{\tilde{\omega}}_t, \qquad \boldsymbol{\tilde{\omega}}_t \sim \mathcal{N}(\boldsymbol{0}, \boldsymbol{Q}_s^{-1}),
\end{equation}
for $t = 1, \dots, T$, with $\boldsymbol{\xi}_1 \sim \mathcal{N}(\boldsymbol{0}, (1 - \rho^2)^{-1} \boldsymbol{Q}_s^{-1})$. The joint distribution of the stacked latent process $\boldsymbol{\xi} = (\boldsymbol{\xi}_1^\top, \dots, \boldsymbol{\xi}_T^\top)^\top$ is thus:
\begin{equation*}
\boldsymbol{\xi} \sim \mathcal{N}(\boldsymbol{0}, \boldsymbol{Q}^{-1}),
\end{equation*}
where $\boldsymbol{Q} = \boldsymbol{Q}_T \otimes \boldsymbol{Q}_s$, and $\boldsymbol{Q}_T$ is the $T$-dimensional AR(1) temporal precision matrix (see \citet{lindgren2011explicit} and \citet{cameletti2013spatio} for details).
Let $\mathbf{y}_t := \mathbf{y}(\cdot, t)$, $\boldsymbol{\eta}_t := \boldsymbol{\eta}(\cdot, t)$, and $\mathbf{z}_t := \mathbf{z}(\cdot, t)$; the linear predictor becomes:
\begin{equation*}
\boldsymbol{\eta}_t = \mathbf{z}_t \boldsymbol{\beta} + \mathbf{A} \boldsymbol{\xi}_t,
\end{equation*}
so the data model (likelihood) given in Equation \eqref{a} becomes:
\begin{equation}\label{lik}
\mathbf{y}_t = \mathbf{z}_t \boldsymbol{\beta} + \mathbf{A} \boldsymbol{\xi}_t + \boldsymbol{\epsilon}_t, \qquad \boldsymbol{\epsilon}_t \sim \mathcal{N}(\boldsymbol{0}, \sigma_\epsilon^2 \mathbf{I}_n),
\end{equation}
where $\mathbf{A}$ is a sparse $n \times m$ matrix projecting the GMRF $\boldsymbol{\xi}_t$ onto the observation locations.

The model defined by Equations \eqref{ar} and \eqref{lik}   is a latent Gaussian model and is estimated using INLA. Denote $\mathbf{x} = (\boldsymbol{\xi}, \boldsymbol{\beta})$ as the latent field with independent prior components. The prior for $\boldsymbol{\xi}$ is GMRF, while vague Gaussian priors were used for all components of $\boldsymbol{\beta}$ to reflect minimum prior knowledge. The joint prior is then given as
\begin{equation*}
\pi(\mathbf{x} \mid \boldsymbol{\theta}) \sim \mathcal{N}(\mathbf{0}, \mathbf{Q}(\boldsymbol{\theta}_1)),
\end{equation*}
where $\boldsymbol{\theta}_1 = (\kappa, \rho, \sigma_\omega^2)$ are the latent model hyperparameters. Observations $\mathbf{y}$ are conditionally independent Gaussian given $\mathbf{x}$ and $\boldsymbol{\theta}_2 = \sigma_\epsilon^2$. Letting $\boldsymbol{\theta} = (\boldsymbol{\theta}_1, \boldsymbol{\theta}_2)$, the joint posterior becomes:
\begin{equation}\label{post}
\pi(\mathbf{x}, \boldsymbol{\theta} \mid \mathbf{y}) \propto \prod_{t=1}^T \pi(\mathbf{y}_t \mid \mathbf{x}, \boldsymbol{\theta}) \pi(\mathbf{x} \mid \boldsymbol{\theta}) \pi(\boldsymbol{\theta}),
\end{equation}
where $\pi(\mathbf{y}_t \mid \mathbf{x}, \boldsymbol{\theta}) \sim \mathcal{N}(\mathbf{z}_t \boldsymbol{\beta} + \mathbf{A} \boldsymbol{\xi}_t, \sigma_\epsilon^2 \mathbf{I}_n)$, and $\pi(\boldsymbol{\theta})$ is the prior for hyperparameters. 
The models are fully specified once prior distributions are set for the hyperparameters $\boldsymbol{\theta}$. In the SPDE approach, the smoothness parameter is fixed at $\nu = 1$. For the spatial range $\rho$ and marginal standard deviation $\sigma_\omega$, we adopted the joint Penalized Complexity (PC) priors proposed by \citet{fuglstad2019constructing}, defined as $\text{Prob}(\rho < 600) = 0.5$ and $\text{Prob}(\sigma_\omega > 3) = 0.01$, reflecting beliefs in broad spatial dependence and moderate variability. For the temporal autocorrelation parameter $\phi$, we used the default PC prior from \citet{sorbye2017penalised}, $\text{Prob}(\phi > 0) = 0.9$, which shrinks toward weak dependence while remaining flexible. These priors regularize model complexity and guard against overfitting. Sensitivity analyses using alternative prior settings and mesh resolutions confirmed the robustness of posterior estimates.
Using this framework, the marginal posterior distributions $\pi(x_i \mid \mathbf{y})$ and $\pi(\theta_j \mid \mathbf{y})$ for $i = 1, \dots, T + p$ and $j = 1, \dots, 4$ are approximated via the INLA algorithm \cite{rue2009approximate}. Inference was performed using the INLA-SPDE approach \cite{bakka2018spatial, lindgren2015bayesian}, which is significantly more computationally efficient than MCMC methods traditionally used in Bayesian inference.

\subsection{Model validation and diagnosis}
To evaluate model fit and predictive performance, we used several Bayesian and classical diagnostics. The Deviance Information Criterion (DIC) \citep{spiegelhalter2002bayesian} and the Watanabe-Akaike Information Criterion (WAIC) \citep{watanabe2013widely} were used for model selection, where lower values indicate better fit with appropriate model complexity. For predictive assessment, we employed the conditional predictive ordinate (CPO) \citep{pettit1990conditional} and the predictive integral transform (PIT) \citep{marshall2003approximate}. CPO values, computed using the logarithmic score (LCPO) \citep{malgorzataroos2011}, measure out-of-sample prediction accuracy, while uniform PIT distributions suggest good calibration. Due to potential numerical instability in INLA, care was taken when interpreting these values. The dataset (2012–2022) was split into training (2012–2021) and validation (2022) periods. Predictive performance was evaluated using the correlation coefficient ($r$), root mean square error (RMSE), and mean absolute error (MAE). High $r$ and low RMSE/MAE indicate better predictive accuracy. 
\begin{equation*}
r = \frac{\sum_{t,i}(y(s_i,t) - \bar{y})(\hat{y}(s_i,t) - \bar{\hat{y}})}{\sqrt{\sum_{t,i}(y(s_i,t) - \bar{y})^2} \sqrt{\sum_{t,i}(\hat{y}(s_i,t) - \bar{\hat{y}})^2}}, \quad
\mathrm{RMSE} = \sqrt{\frac{1}{n} \sum_{t,i}(y(s_i,t) - \hat{y}(s_i,t))^2},
\end{equation*}
\begin{equation*}
\mathrm{MAE} = \frac{1}{n} \sum_{t,i} \left| y(s_i,t) - \hat{y}(s_i,t) \right|.
\end{equation*}
Together, these metrics provide a comprehensive assessment of model adequacy and predictive validity.


\section{Results}\label{result}
In this section, we first compare how well the Bayesian spatiotemporal models work and choose the one that can predict the future the best (Section \ref{mod_co}). Then, we summarize  posterior estimates for both fixed and random effects (Section \ref{mod_sum}). 
Note that since the explanatory variables are measured on different scales, to avoid numerical problems, each predictor is standardized to have a mean of zero and a unit standard deviation.


\subsection{Model comparison}\label{mod_co}
\begin{figure}[H]
\centering
\includegraphics[width=1\linewidth]{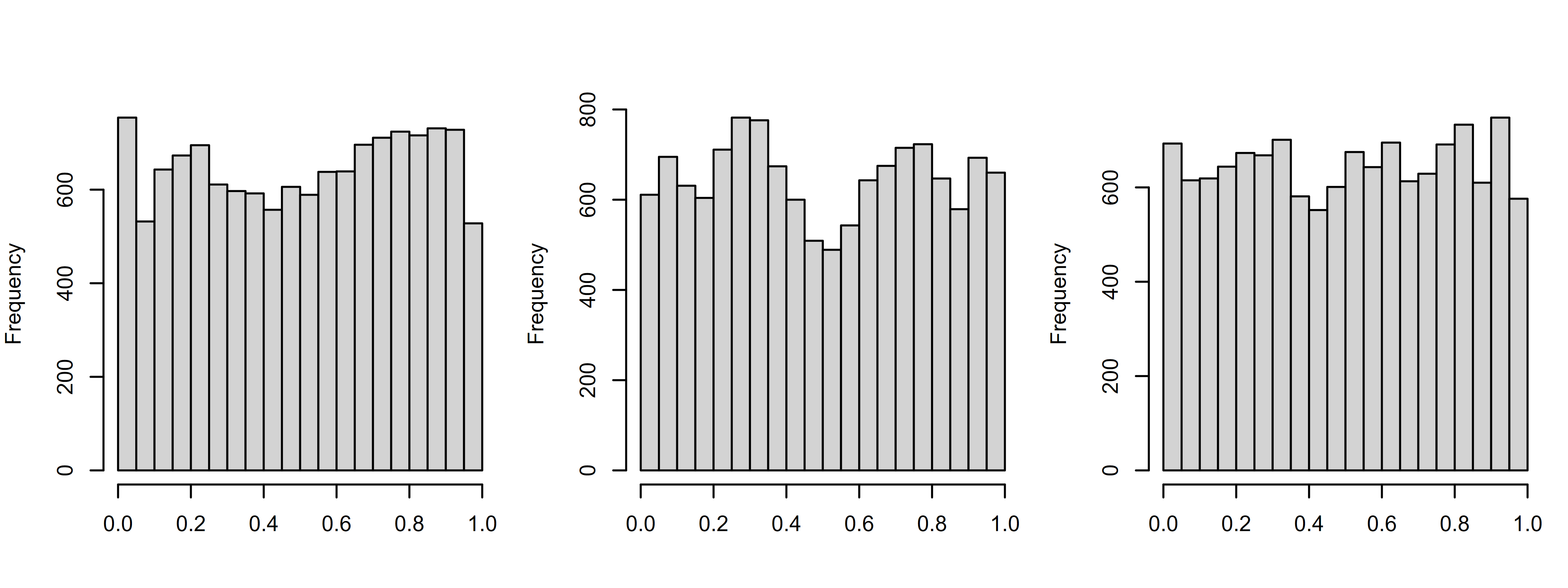}
\caption{The picture shows the predictive integral transform (PIT) plots for Model 1 (left), Model 2 (middle), and Model 3 (right), which are based on Equations \eqref{c}, \eqref{5}, and \eqref{1}. A uniform distribution pattern is preferable.}
\label{pit}
\end{figure}

\begin{figure}[H]
\centering
\includegraphics[width=1\linewidth]{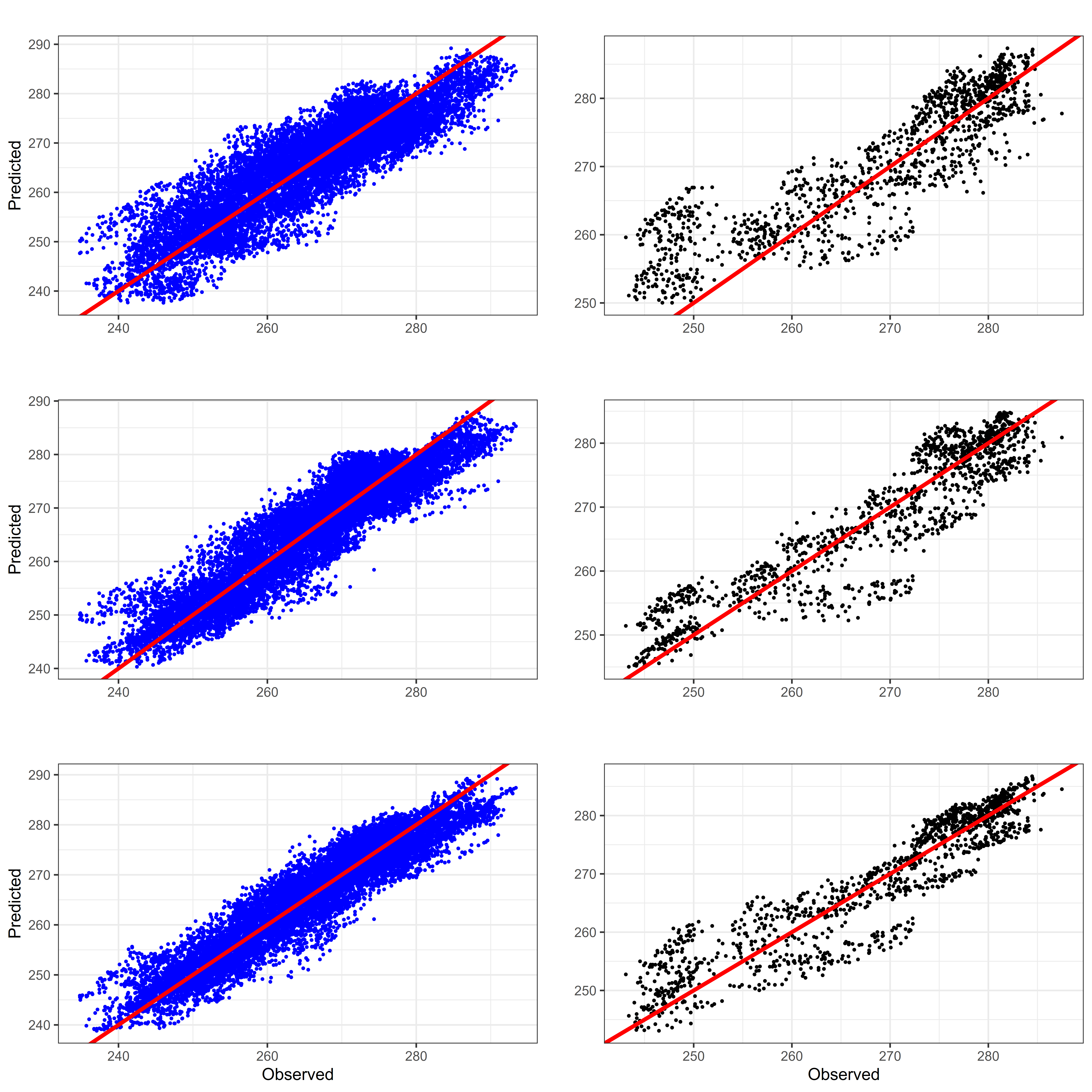}
\caption{Visualization of training (left) and validation (right) performance for Model 1 (top), Model 2 (middle), and Model 3 (bottom) specified in Equation \eqref{c}, \eqref{5}, and \eqref{1}, respectively. The scatters distributed along the line with the intercept 0 and the slope 1 denote the better model, and Model 3 is the best among all.}
\label{val}
\end{figure}

Different evaluation criteria are used on the training dataset (DIC, WAIC, PIT, and RMSE) and the validation dataset (PIT, RMSE, and MAE correlation) to see how well the models fit and can predict the TCO value. These can be seen in Table \ref{mod_c1}. Model 3 outperforms in DIC (9451.45), WAIC (9462.09), LCPO (0.365), and correlation (0.94) for the training set and also has smaller values for RMSE (4.445 DU) and a higher value in correlation (0.92) in the validation set. In terms of the PIT plots on the training set (Figure \ref{pit}), which favors the shape of uniform distribution, all models perform better.
\begin{table}[H]
\begin{adjustbox}{width=\textwidth}
\centering 
\begin{tabular}{lcccccccc}
\toprule
& \multicolumn{4}{c}{Training Set (2012-2021)} && \multicolumn{3}{c}{Validation Set (2022)} \\
\cmidrule{2-5} \cmidrule{7-9}
Models & DIC & WAIC & LCPO & Correlations && RMSE & MAE & Correlations \\
\midrule 
Model 1 & 16995.93 & 16996.241 & 0.656 & 0.883 && 5.997 & 4.604 & 0.813
\\
Model 2 & 11823.064 & 11824.348 & 0.456 & 0.924 && 4.664 & 3.556 & 0.902\\
Model 3 & 9451.498 & 9462.095 & 0.365 & 0.940 && 4.445 & 3.454 & 0.921\\
\bottomrule
\end{tabular}
\end{adjustbox}
\caption{Model performance and comparison on the training and validation datasets of Models 1-3 specified in Equations \eqref{c}, \eqref{5}, and \eqref{1}. A preferable model is characterized by higher correlation and lower values in RMSE and MAE.}
\label{mod_c1}
\end{table}
Figure \ref{val} shows the simultaneous visualization of model performance on both training and validation sets, where the x-axis represents the observed values and the y-axis represents the estimated (predicted) values. 
The scatters distributed along the line with intercept 0 and slope 1 mean the estimated (predicted) values are best suited to the observations. Model 3 generally performs better than Model 1 and Model 2 because the scatter points in the sub-figure at the bottom are clustered more closely around the identity line, indicating a stronger alignment between observed and estimated values. Also, Models 1 and 2 tend to give lower estimated (predicted) values when observed values go up, as seen in Figures \ref{val} top and middle. This is in line with the larger RMSE. In summary, from the three models, Model 3 given in Equation \eqref{1} excels in all categories of criteria and gives a satisfactory result in most circumstances (DIC, WAIC, RMSE, correlation, and the visualization plot) for different combinations of predictors.


\subsection{Model summary}\label{mod_sum}
\begin{figure}[H]
\centering
\includegraphics[width=1\linewidth]{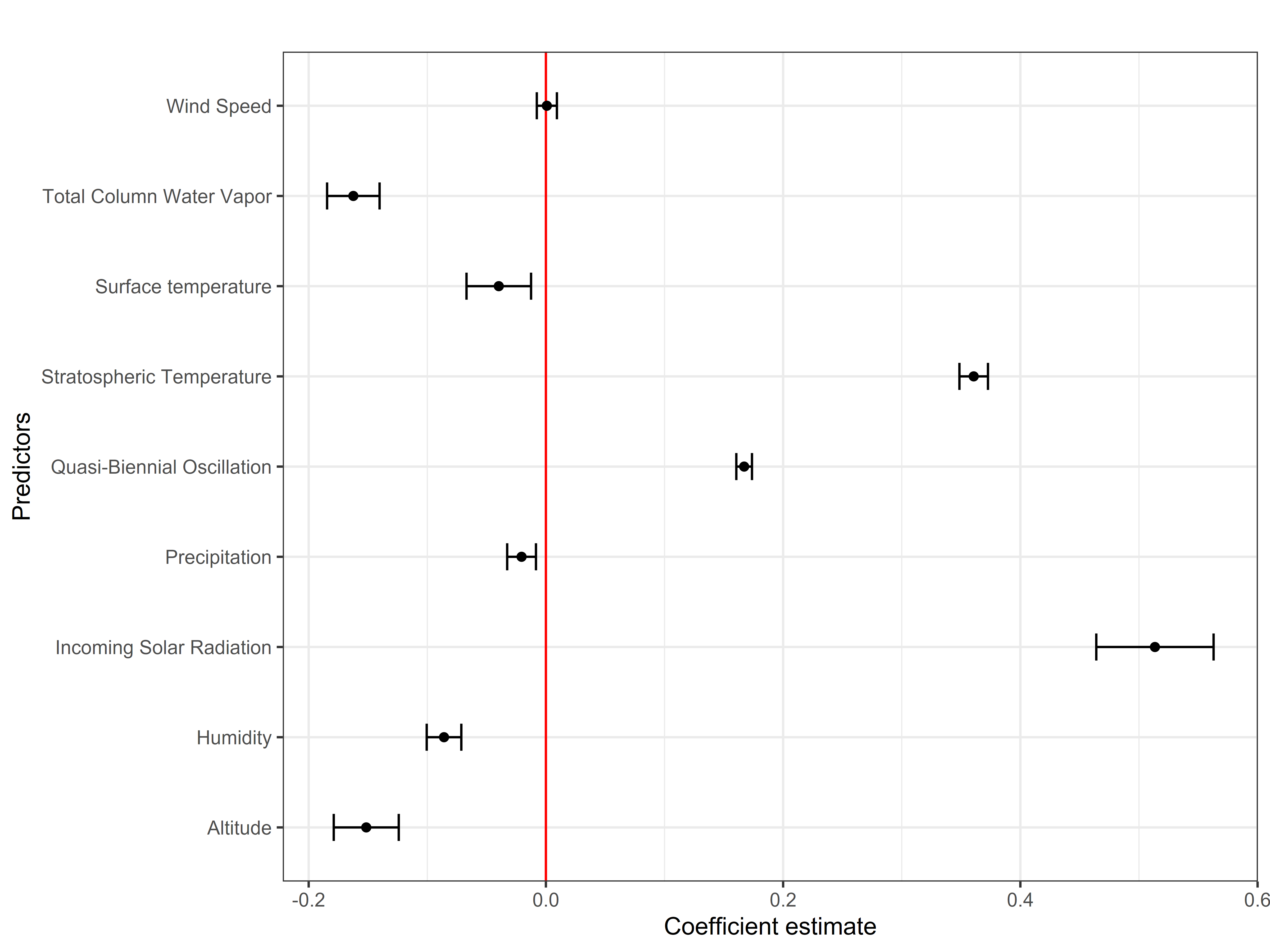} 
\caption{Visualization of the posterior estimates of parameters for the fixed effect}
\label{fix_plt}
\end{figure}

We looked at the posterior estimates of standardized regression coefficients from the fitted spatiotemporal Bayesian model (Equation \eqref{1}) to figure out how much meteorological, climatological, and geographic factors affect changes in total column ozone (TCO) over Ethiopia. After taking into account effects on space, time, and interactions, these coefficients show the change in TCO (in standard deviation units) that happens when each predictor moves by one standard deviation (SD). This standardization facilitates direct comparison of predictor effects across differing scales and units. The direction, magnitude, and uncertainty of these associations identify the main drivers from the predictors of TCO variability in the region. Full posterior summaries (means, SDs, and 95\% credible intervals) for all parameters are provided in Table \ref{fix_eff}, with visualizations in Figure \ref{fix_plt}.
\begin{table}[H]
\centering
\begin{adjustbox}{width=\textwidth}
\begin{tabular}{l|cccc}
\hline
\textbf{Covariates}& Mean & Stdev & 0.025 quantile  & 0.95 quantile   \\ \hline
Surface temperature & -0.040 & 0.014  & -0.067 &  -0.013 \\ 
Precipitation &-0.021 & 0.006 & -0.033  & -0.008 \\ 
Wind Speed & 0.001 & 0.004 & -0.008 &  0.009  \\
Humidity & -0.086 & 0.007 & -0.100 & -0.071 \\ 
Stratospheric temperature & 0.361 & 0.006 & 0.349  &  0.373  \\ 
Total column water vapor & -0.162 & 0.011 & -0.184  &  -0.140 \\
Incoming solar radiation & 0.513 & 0.025 & 0.464  &   0.563 \\   
Quasi-Biennial Oscillation & 0.167 & 0.003 & 0.161  &  0.174 \\  
Altitude & -0.151 & 0.014 & -0.179  & -0.124 \\ 
\hline
\end{tabular}
\end{adjustbox}
\caption{Posterior estimates (mean, standard deviation, and quantiles) of the coefficients of the fixed effects from Model 3.}
\label{fix_eff}
\end{table}
The intercept, which shows the expected TCO when all predictors are at their mean values (standardized to 0), is thought to be 0.199 SD, and its 95\% confidence interval is between 0.019 and 0.380. Translating this to absolute TCO values (mean = 266.1 DU, SD = 11.5 DU), this corresponds to 268.4 DU $(266.1 + 0.199\times 11.5)$ under average conditions. The range is from 266.3 DU (0.019 SD) to 270.5 DU (0.380 SD), which shows that Ethiopia's TCO is slightly higher than the study period mean even when climatological predictor states are neutral. This baseline enrichment is probably because Ethiopia is near the equator in the ozone production belt, which has its own unique dynamics for how the stratosphere and troposphere interact with each other. The precision of this estimate (Stdev = 0.092), as shown in Table \ref{fix_eff}, underscores its reliability as a reference for interpreting covariate effects, which are deviations from this ozone-retentive baseline.
Detailed interpretations of each covariate’s effect on TCO, grounded in atmospheric processes and Ethiopian climatic context, are presented in the subsections below.

\subsubsection{Effect of incoming solar radiation on TCO}
The most positive standardized effect is seen in incoming solar radiation (ISR), with a posterior mean of 0.513 and a 95\% credible interval (CI) of (0.464, 0.563). This implies that a one standard deviation (SD) increase in ISR corresponds to a 0.513 SD increase in TCO over Ethiopia, controlling for other variables. The narrow CI indicates high confidence in this robust photochemical relationship. According to photochemical theory \cite{chapman1930xxxv}, ozone is made through the Chapman cycle when strong equatorial ISR, especially UV radiation, breaks down oxygen molecules (O\textsubscript{2}) into ozone (O\textsubscript{3}). Ethiopia's equatorial position and high elevation jointly enhance UV penetration into the stratosphere, increasing local photochemical efficiency. Yet, this local production signal exists alongside the equatorial ozone paradox: although the tropics are a major ozone source, total column ozone is lower than at mid-latitudes due to poleward transport by the Brewer-Dobson circulation \cite{butchart2014brewer}. The positive ISR coefficient suggests that, despite this export, local photochemistry remains a dominant control on TCO variability in Ethiopia. These results underscore Ethiopia’s dual role as both an ozone-producing region and a net exporter. It also highlights practical implications for UV exposure, as high ISR and elevation can elevate surface UV levels even in the context of modest TCO values. More broadly, the findings stress the importance of separating local production from large-scale transport when studying ozone dynamics in the tropics, especially under varying climate conditions.
\subsubsection{Effects of stratospheric temperature on TCO}
The temperature of the stratosphere at 30 hPa was strongly linked to TCO over Ethiopia, with a mean value of 0.361 and a 95\% confidence interval of 0.349 to 0.373. The narrow credible interval indicates high confidence in this relationship, where a 1 SD increase in stratospheric temperature corresponds to a 0.361 SD rise in TCO, keeping other predictors constant. In a chemical sense, warmer temperatures slow down ozone-depleting catalytic cycles, especially those involving NO and ClO \cite{solomon1999stratospheric}. On the other hand, it speeds up the Chapman cycle \cite{chapman1930xxxv}, which makes more ozone by speeding up the reactions of O + O\textsubscript{2}$\rightarrow$ O\textsubscript{3}. Dynamically, elevated temperatures strengthen Brewer-Dobson circulation, redistributing ozone-rich air downward and suppressing tropical upwelling. Ethiopia’s high elevation amplifies this coupling by reducing tropospheric buffering, allowing stratospheric processes to dominate local TCO variability. This finding aligns with studies \cite{randel2007stratospheric, shangguan2019variability} demonstrating lower-stratospheric cooling’s role in tropical ozone retention, particularly during periods of reduced solar heating. However, long-term stratospheric cooling driven by greenhouse gas emissions threatens to counteract these photochemical gains, highlighting the dual role of stratospheric temperature as both a driver of TCO variability and a sentinel for ozone layer resilience.

\subsubsection{Effect of the Quasi-Biennial Oscillation on TCO}
\begin{figure}[ht]
\centering
\includegraphics[width=1\linewidth]{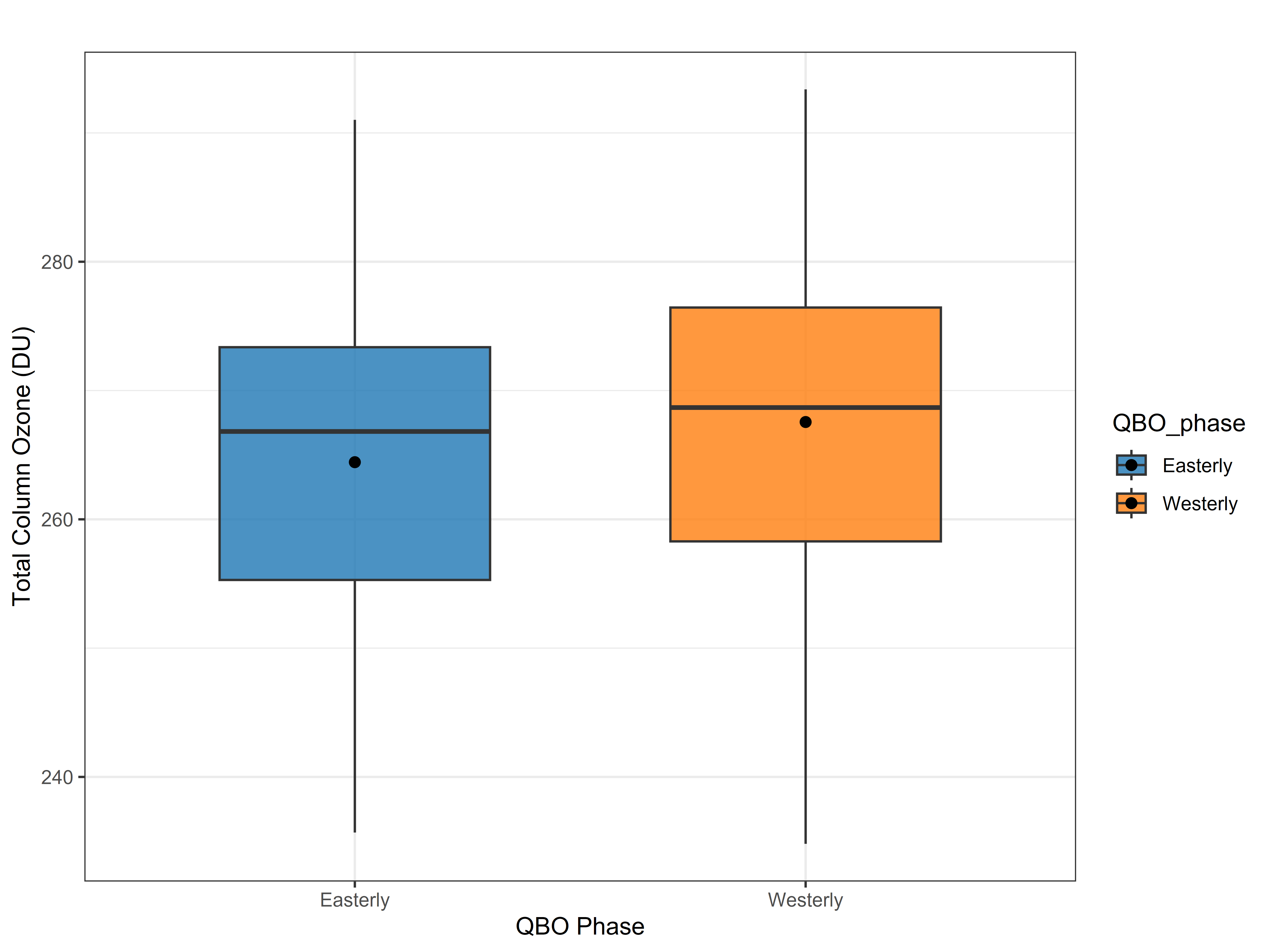}
\caption{Box plots of the distribution of total TCO over Ethiopia during easterly and westerly phases of the QBO from 2012 to 2022. The figure shows that TCO levels are higher during westerly QBO phases (mean = 268 DU) than during easterly phases (mean = 264 DU), consistent with known stratospheric dynamics where westerly winds suppress upwelling of ozone-poor air, promoting ozone accumulation. Points indicate phase-wise mean TCO values.}
\label{qphase}
\end{figure}

The Quasi-Biennial Oscillation (QBO) has been identified as a significant factor influencing TCO variability among climatological predictors. The estimated standardized regression coefficient for the QBO index is 0.167, with a 95\% CI of (0.161, 0.174). This means that a one standard deviation rise in zonal wind at 30 hPa, which means a change toward stronger westerly flow, is equal to a 0.167 standard deviation rise in TCO when all other factors are taken into account. This link fits with what we know about how the tropical stratosphere works. Westerly QBO phases stop tropical upwelling, slow down the movement of ozone-poor air up and down, and then encourage ozone buildup in the lower stratosphere. This result is consistent with the study by \cite{wang2022zonally}, and exploratory data analyses have confirmed the dynamics. The mean TCO is higher during westerly QBO phases (268 DU) than during easterly phases (264 DU), which is in line with the regression direction as depicted in Figure \ref{qphase}.
The results show how important changes in the stratospheric wind are for affecting the changes in ozone levels between years in Ethiopia, which is close to the equator and has strong QBO signals. This empirical pattern supports the assertion that the QBO is a significant factor influencing interannual and seasonal ozone variability in Ethiopia.

\subsubsection{Effect of total column water vapor on TCO}
The estimated standardized regression coefficient for total column water vapor (TCWV) is –0.162 with a 95\% credible interval of (–0.184, –0.140), indicating that a one standard deviation increase in TCWV corresponds to a 0.162 SD decrease in TCO over Ethiopia, holding other predictors constant. This inverse relationship reflects water vapor’s dual contribution to ozone depletion. Chemically, elevated TCWV enhances the abundance of hydroxyl radicals (OH), which drive catalytic ozone loss through HO\textsubscript{x} cycles (e.g., O\textsubscript{3} + OH $\rightarrow$ HO\textsubscript{2} + O\textsubscript{2}) \cite{solomon1999stratospheric}. Radiatively, water vapor’s greenhouse effect promotes stratospheric cooling, which slows ozone production rates \cite{solomon1999stratospheric}. In Ethiopia, TCWV peaks during the June–September rainy season, particularly over the humid southern highlands, where intense convection transports moisture into the lower stratosphere. This amplifies local ozone-depleting processes even as large-scale circulation like the Brewer-Dobson mechanism redistributes ozone poleward \cite{butchart2014brewer}. These findings underscore the importance of hydrological variability in shaping tropical ozone chemistry and the need to accurately model seasonal moisture transport when analyzing TCO dynamics in regions with seasonal rainfall patterns.

\subsubsection{Effect of the altitude on TCO}
The estimated standardized regression coefficient for altitude is –0.151 with a 95\% credible interval of (–0.179, –0.124), indicating that a one standard deviation increase in elevation corresponds to a 0.151 SD decrease in TCO over Ethiopia, holding other variables constant. This inverse relationship suggests that high-altitude regions, such as the Ethiopian Highlands, exhibit systematically lower TCO levels. Several mechanisms may explain this effect: elevated terrain reduces tropospheric thickness, enhancing stratosphere–troposphere exchange \cite{ball2019inconsistencies} and facilitating the downward intrusion of ozone-poor air. Additionally, stronger UV radiation at higher elevations can increase ozone photolysis (O\textsubscript{3} + UV $\rightarrow$ O\textsubscript{2} + O) \cite{dewan2022tropospheric}, while enhanced wind shear may accelerate ozone export via the Brewer-Dobson circulation \cite{cohen2014drives}. Despite being closer to the ozone-rich stratospheric layers, these regions appear to experience net ozone dilution. Similar altitude-related TCO reductions have been reported over other high-elevation tropical regions, such as the Tibetan Plateau, where low TCO has been linked to topography and seasonal tropopause dynamics \cite{kuttippurath2023trends}. The findings underscore elevated UV exposure risks in Ethiopian cities like Addis Ababa and Gondar due to the combined effects of high altitude and reduced ozone shielding.

\subsubsection{Effect of surface temperature on TCO}
Surface temperature shows a significant negative association with TCO over Ethiopia, with a standardized regression coefficient of –0.040 and a 95\% credible interval of (–0.067, –0.013). This means that a one standard deviation increase in surface temperature corresponds to a 0.04 SD decrease in TCO, holding other variables constant. Elevated surface temperatures promote vertical mixing, allowing ozone-poor air and pollutants such as industrial emissions and aerosols from biomass burning to rise into the lower stratosphere \cite{ball2019inconsistencies}. These substances accelerate ozone loss through photochemical reactions involving chlorine and bromine compounds. Surface warming also increases infrared radiation emitted from the ground, which cools the stratosphere and reduces ozone production efficiency \cite{solomon1999stratospheric, ball2019inconsistencies}. In Ethiopia, this effect is especially pronounced in lowland areas like Afar, where extreme heat and convection during the dry season and the months leading up to the rainy season support upward transport of ozone-depleting substances. While the seasonal rains help counteract this by cooling the surface and increasing cloud cover, the findings suggest that ongoing warming could gradually intensify ozone depletion and increase surface UV exposure risks.

\subsubsection{Effect of precipitation on TCO}
Monthly precipitation showed a statistically significant negative relationship with TCO over Ethiopia, with a standardized regression coefficient of –0.021 and a 95\% credible interval of (–0.033, –0.008). This suggests that a one standard deviation increase in precipitation corresponds to a 0.021 SD decrease in TCO, holding other factors constant. The inverse relationship likely reflects the combined effects of increased cloud cover, convective activity, and wet scavenging \cite{gryspeerdt2015wet}. Enhanced cloudiness during wetter months reduces incoming UV radiation, suppressing photolytic ozone production in the stratosphere. Simultaneously, deep convection activity, especially during the rainy season of June–September, can loft ozone-depleting substances into the lower stratosphere, accelerating ozone loss through catalytic cycles \cite{solomon1999stratospheric}. Furthermore, precipitation may promote the wet removal of ozone precursors like NO\textsubscript{x} via cloud microphysics \cite{houze2014cloud}. These processes together contribute to lower TCO during the rainy season, particularly over Ethiopia’s western highlands.

\subsubsection{Effect of humidity on TCO}
The estimated standardized regression coefficient for upper tropospheric humidity is –0.086 with a 95\% credible interval of (–0.100, –0.071), indicating that a one standard deviation increase in humidity is associated with a 0.086 SD decrease in TCO over Ethiopia, holding other predictors constant. This negative association suggests that increased humidity contributes directly to ozone depletion through several mechanisms. Higher humidity enhances the hydrolysis of nitrogen oxides on cloud particles, reducing their availability for catalytic ozone production \cite{dewan2022tropospheric}. It also promotes the formation of hydroxyl radicals that accelerate ozone loss \cite{solomon1999stratospheric}. During the main rainy season in southwestern Ethiopia, elevated humidity leads to infrared radiation that cools the stratosphere and supports the dissolution and activation of halogen compounds such as bromine monoxide from biomass burning. These combined effects highlight the significant role of humidity-driven processes in modulating ozone concentrations in the upper troposphere and lower stratosphere.

\subsubsection{Effect of wind speed on TCO}
Moreover, the estimated coefficient for the V-component wind speed at 30 hPa is 0.001 with a 95\% CI (–0.008, 0.009), indicating a very weak and statistically insignificant relationship with TCO over Ethiopia. The near-zero coefficient implies that, in the context of this model, variability in meridional wind at 30 hPa does not have a consistent or strong influence on TCO over Ethiopia during the study period. While meridional winds can facilitate cross-latitudinal ozone transport, bringing ozone-rich air from mid-latitudes or drawing in ozone-poor air from the tropics, their net impact over this specific region and time frame appears minimal. This could reflect the complex interplay of competing transport mechanisms or the dominant influence of other predictors like stratospheric temperature or QBO in modulating stratospheric ozone.


\subsection*{Hyperparameter estimates}
\begin{figure}[H]
\centering
\includegraphics[width=1\linewidth]{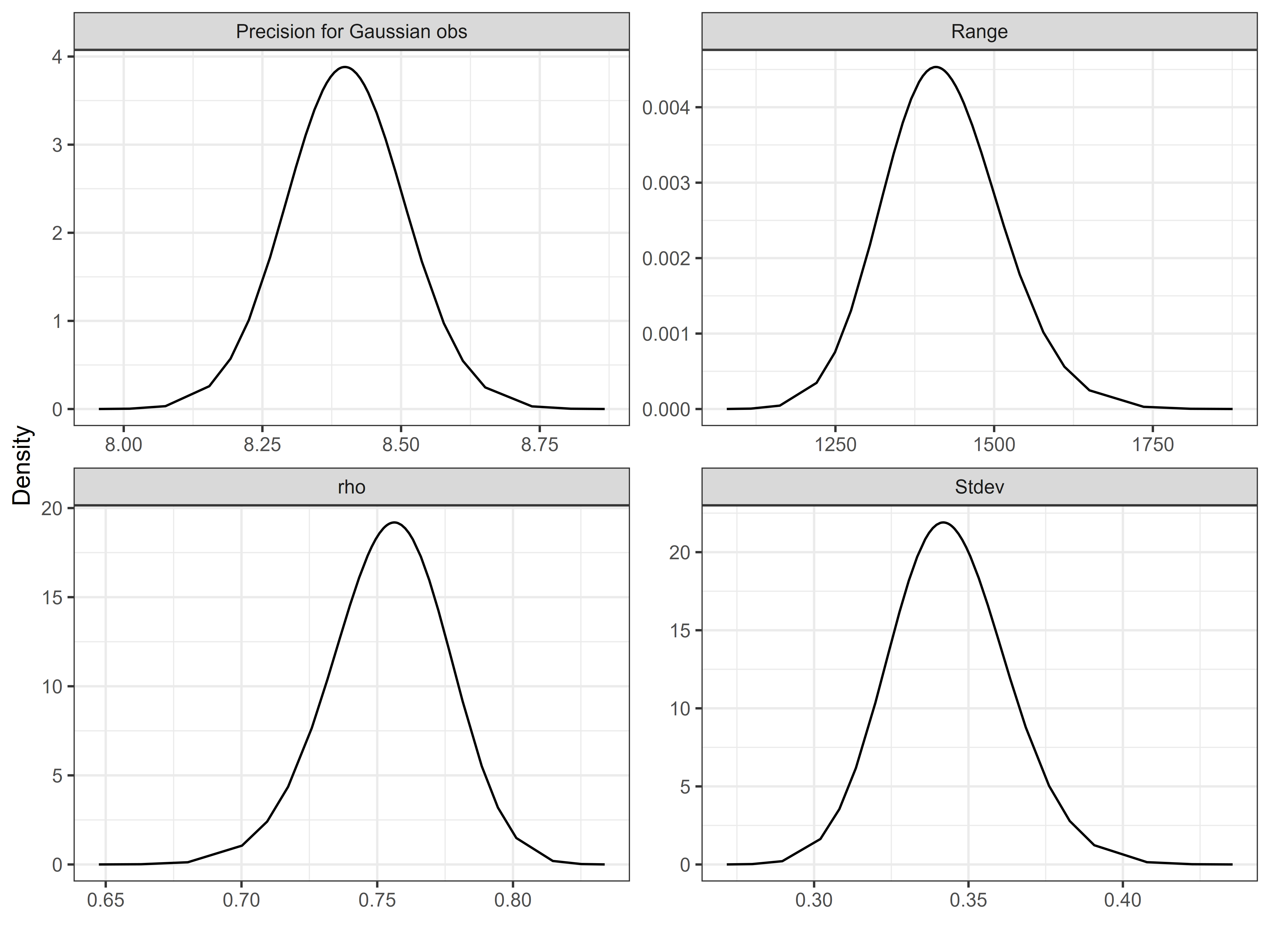} 
\caption{Visualization of the posterior estimates of the hyperparameters}
\label{hyper}
\end{figure}

The estimated posterior means, standard deviations, and 95\% credible intervals of model hyperparameters $(\tau_\epsilon, \rho, \sigma_\omega, \phi)$ for the latent spatiotemporal process are presented in Table \ref{hyper_par} and its visualization in Figure \ref{hyper}. The spatiotemporal random effects are a mechanism of capturing the unexplained patterns not explained by other model components in the model that presents spatial dependence, and in this way they serve to incorporate the effects of covariates and other effects that bring spatial dependence \cite{laurini2019spatio}. 
\begin{table}[H]
\centering
\begin{adjustbox}{width=\textwidth}
\begin{tabular}{l|cccc}
\hline
\textbf{Parameters}& Mean & Stdev & 0.025 quantile & 0.975 quantile   \\ \hline
Precision ($\tau_\epsilon$) for the observation & 8.41 &  0.107 & 8.19 &   8.61 \\ 
Range ($\rho$) & 1419.84 & 91.83 & 1249.26 &  1610.67 \\ 
Stdev ($\sigma_\omega$) & 0.344 & 0.012 & 0.308 &  0.383 \\ 
Autocorrelation ($\phi$) & 0.75 & 0.022 & 0.71 &  0.81 \\
\hline
\end{tabular}
\end{adjustbox}
\caption{Posterior estimates of the mean, standard deviation, and quantiles of the parameters in Model 3. The precision ($\tau_\epsilon$) for Gaussian distribution, the range parameter, $\rho$ and the standard deviation ($\sigma_\omega$) introduced in the Mat\'ern covariance function and the autocorrelation coefficient $\phi$.}
\label{hyper_par}
\end{table}
The precision ($\tau_\epsilon$) for the Gaussian observation is 8.41 with a 95\% credible interval of (8.19, 8.61). This corresponds to a measurement error variance of 0.119 and a standard deviation ($\sigma_\epsilon$) of 0.345. This indicates a relatively low level of observational noise, suggesting that the satellite-derived TCO data are of reasonably high quality.
The estimated marginal standard deviation ($\sigma_\omega$) for the latent spatiotemporal random effect is 0.344 with a 95\% credible interval of (0.308, 0.383), reflecting moderate variability in underlying environmental dynamics not captured by the fixed effects. This latent component captures structured spatiotemporal variation, such as regional ozone transport patterns and influences from unobserved covariates.
We observe that the spatial component shows a similar variability with the measurement error. While the measurement error is relatively low, the near equality between $\sigma_\epsilon$ and $\sigma_\omega$ implies that residual variance is almost evenly divided between observational noise and unaccounted spatiotemporal structure. This balanced partitioning highlights opportunities to enhance model performance: reducing measurement error (via improved TCO retrieval techniques) and lowering latent variability (through finer-scale or more relevant covariates) could significantly improve both explanatory power and predictive accuracy.

In our analysis, the estimated spatial range ($\rho$) has a posterior mean of 1419.84 km with a 95\% credible interval of (1249.26, 1610.67). Although this effective spatial range may appear large at first glance, it is scientifically plausible given the nature of TCO and the dominant atmospheric processes involved. Since TCO represents the integrated ozone concentration over the entire vertical atmospheric column, it is inherently smooth and primarily influenced by large-scale stratospheric circulations and photochemical reactions rather than localized phenomena exhibiting small-scale variability. Consequently, the spatial correlation in the latent process decays slowly over distance, with ranges on the order of 1400 km or more consistent with the physical behavior of TCO. This finding aligns well with previous studies employing the SPDE approach within a hierarchical Bayesian framework to model similar environmental fields, where smooth spatial processes produce long-range dependencies (e.g., \citeauthor{cameletti2013spatio}). In the specific geographic context of Ethiopia and based on the spatial patterns observed in Figure \ref{spa_plot}, the estimated posterior spatial range effectively captures the dominant atmospheric influences governing ozone distribution across the region (for example, QBO, stratospheric temperature, and incoming solar radiation).

The estimated posterior mean of $\phi$, the AR(1) coefficient for the latent spatiotemporal process is 0.75 (CI: (0.71, 0.81)), indicating strong temporal persistence. This means the underlying state evolves gradually over time, with values at consecutive time points highly correlated. 

\begin{figure}[H]
\centering
\includegraphics[width=1\linewidth]{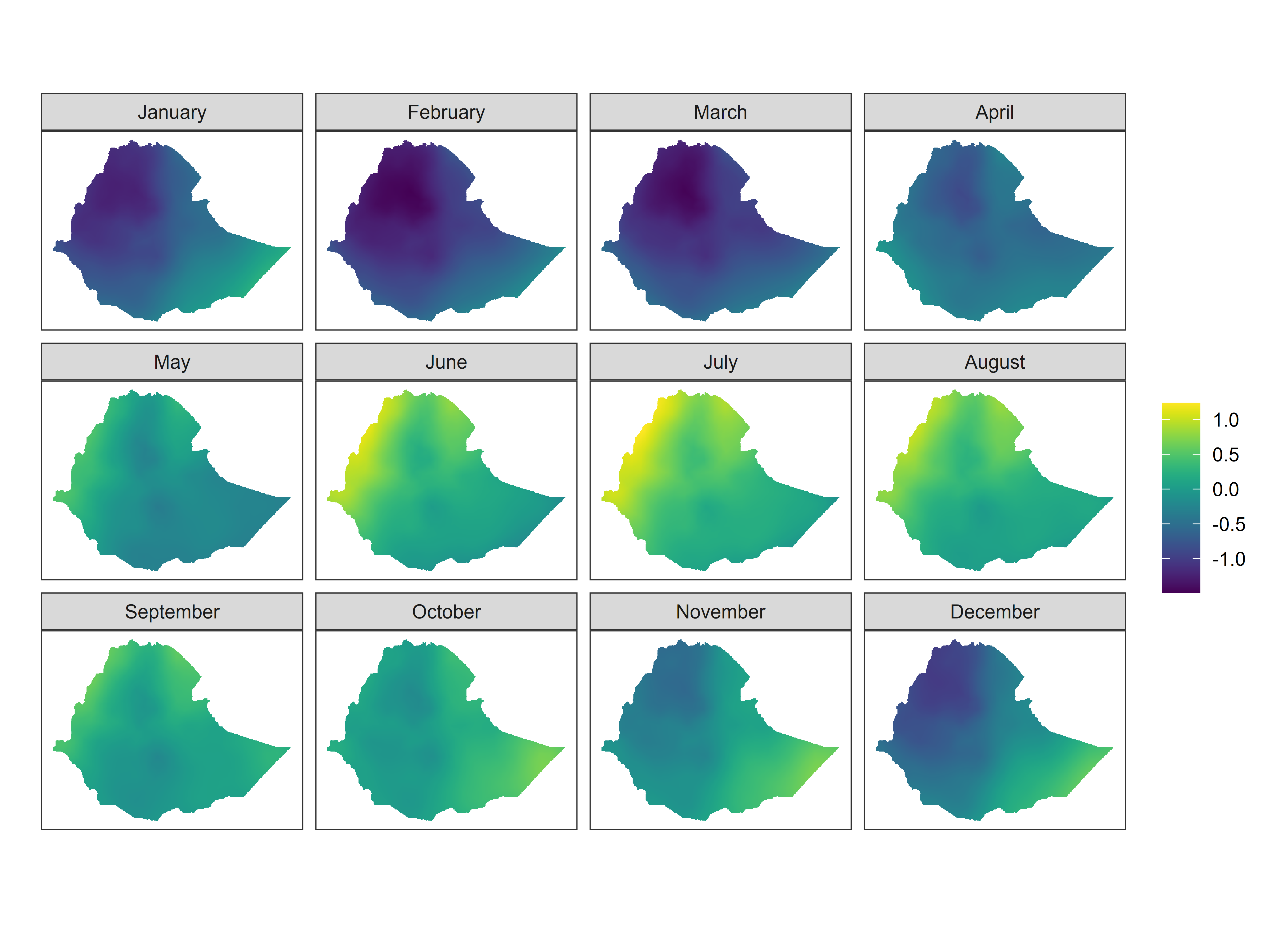} 
\caption{Monthly posterior mean of the latent spatiotemporal process $\xi(s,t)$ estimated using Model 3. The plots show seasonal patterns modeled with a cyclic spatiotemporal interaction structure, where the effect for each calendar month is assumed to remain the same across all years in the study period.}
\label{mean_grf}
\end{figure}

\begin{figure}[H]
\centering
\includegraphics[width=1\linewidth]{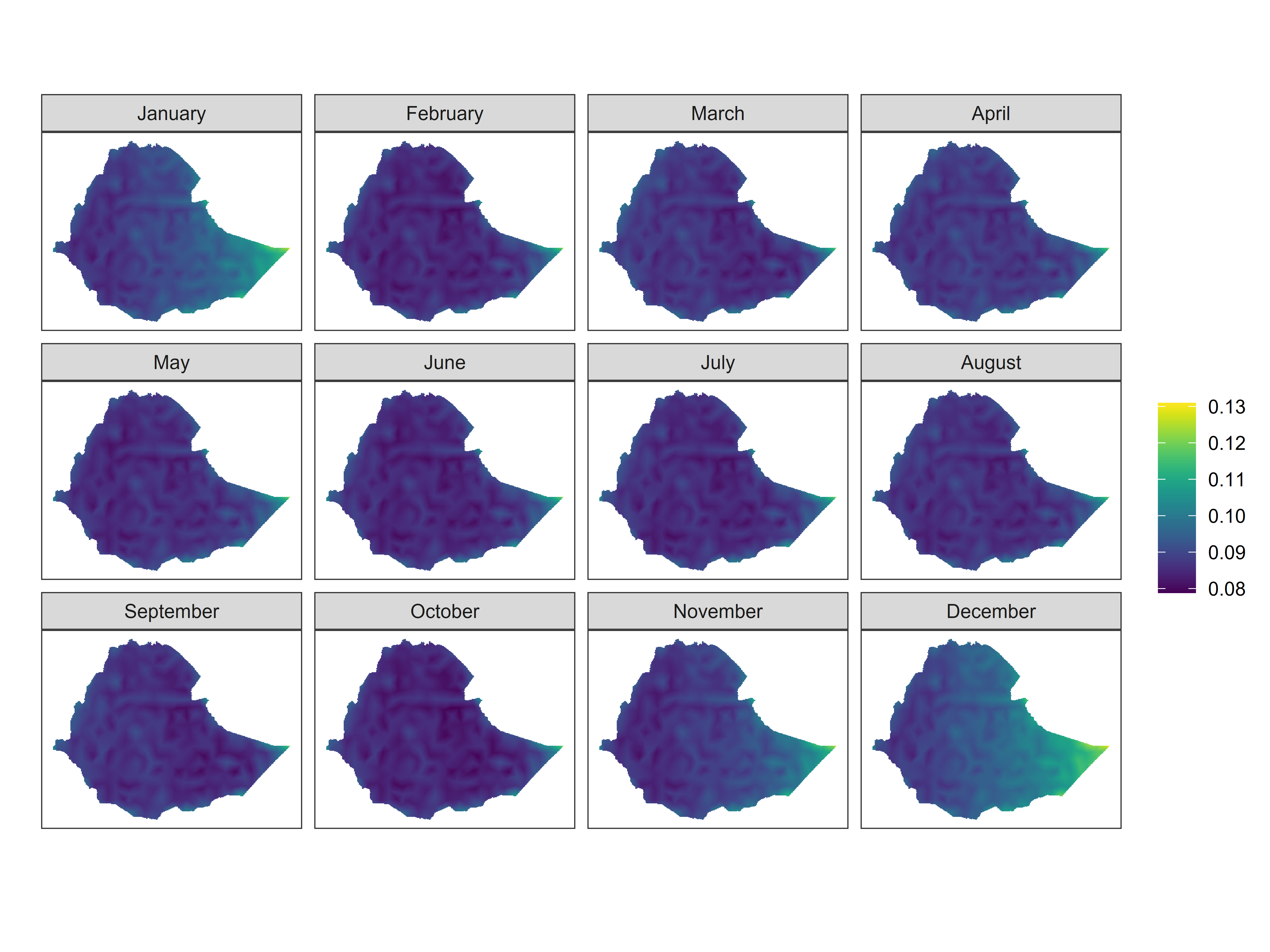} 
\caption{Monthly posterior standard deviation of the latent spatiotemporal process $\xi(s,t)$ estimated using Model 3. The plots show the uncertainty associated with the seasonal patterns modeled using a cyclic spatiotemporal structure, where the effect for each calendar month is assumed to be the same across all years in the study period.}
\label{sd_grf}
\end{figure}

The posterior mean and standard deviation plots of the latent spatiotemporal process $\xi(s,t)$, shown in Figures \ref{mean_grf} and \ref{sd_grf} respectively reveal spatially smooth patterns that evolve gradually over time. These effects are consistent with known global TCO behavior and reflect strong spatial dependence and temporal persistence suggested by the estimated model parameters. The latent process captures residual spatiotemporal variation not explained by the fixed effects, highlighting the model’s ability to represent slow-changing dynamics in ozone concentrations. The monthly posterior mean plots reflect seasonal patterns modeled using a cyclic temporal structure, where each calendar month is assumed to have the same effect across all years of the study period.
This recurring seasonal variability may be influenced by changes in solar radiation \cite{Nassif_2020}, the annual migration of the Intertropical Convergence Zone (ITCZ) \cite{nade2020intra}, large-scale stratospheric dynamics such as the Brewer–Dobson circulation \cite{butchart2014brewer}, and regional meteorological factors.
%
%

The seasonal dynamics of the latent field are particularly evident. During January to March and again in December, negative residuals are concentrated over northern and central Ethiopia, indicating ozone levels lower than expected after accounting for covariates. These anomalies may reflect either reduced ozone transport into Ethiopia or increased export due to a strengthened Brewer-Dobson circulation, both of which are common during boreal winter. In contrast, from June to August, the residuals shift markedly, with positive anomalies emerging across the north, consistent with enhanced photochemical ozone production and altered stratospheric circulation during the summer season. The transitional months of April, May, September, and October display weaker or more spatially diffuse anomalies, reflecting atmospheric shifts between seasonal regimes.
%
Importantly, these spatial and temporal patterns reinforce that the latent process is not only modeling stochastic noise but is capturing physically meaningful environmental signals. These may include sub-grid scale meteorological variability, topographic influences, or other ozone-modulating processes that are not explicitly represented among the covariates. The distinct spatiotemporal structure observed in the Figure \ref{mean_grf}, underscores the importance of including latent random effects in ozone models. Furthermore, the patterns suggest that incorporating finer-resolution or more physically relevant covariates, such as tropopause dynamics, regional wind patterns, or vertical ozone profiles, may reduce residual variance and enhance the explanatory power of the model.

\subsection{Spatial prediction}
\begin{figure}[H]
\centering
\includegraphics[width=1\linewidth]{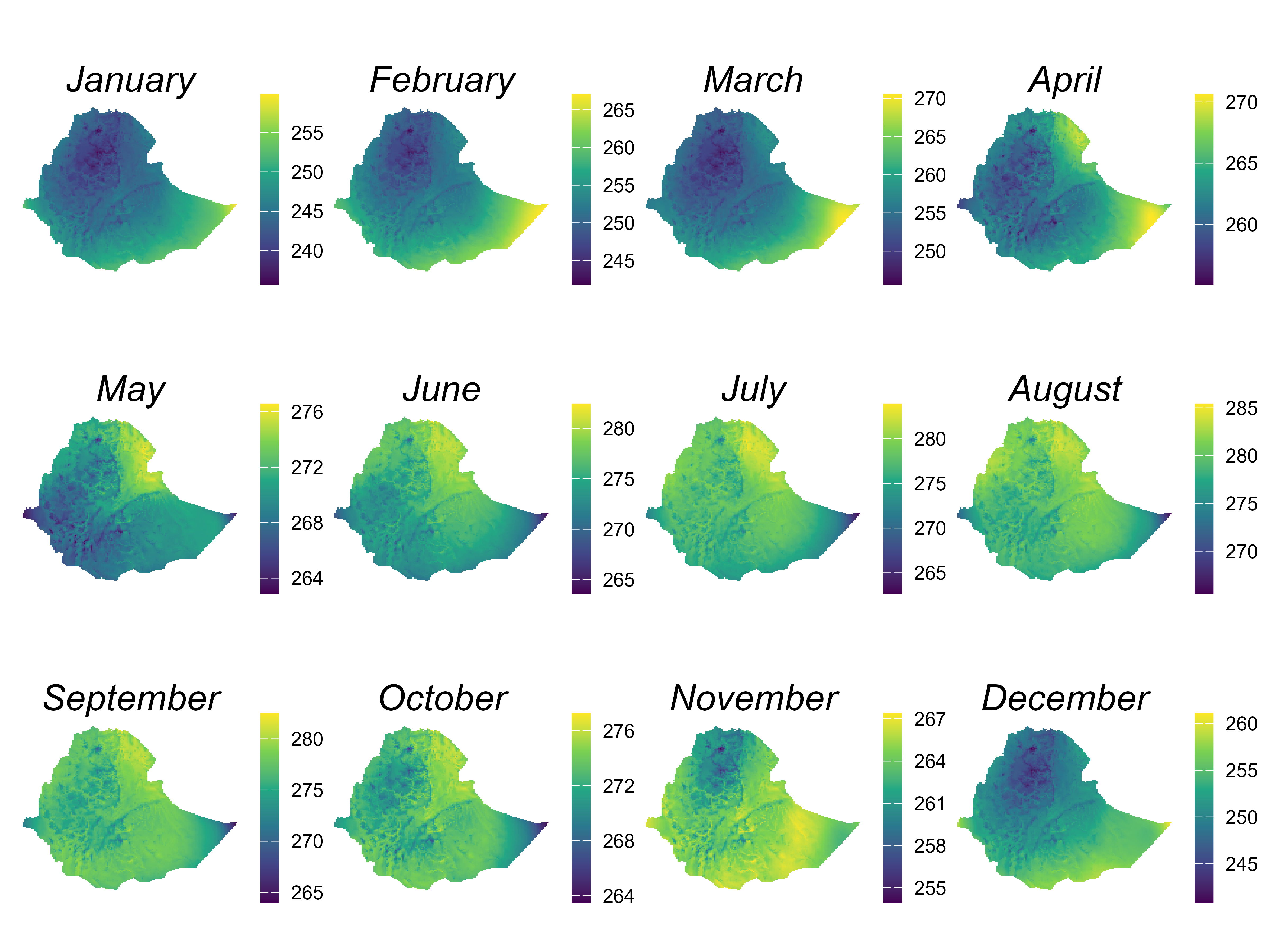} 
\caption{Predictive posterior mean. The figure shows the Bayesian estimation of the predicted mean of the TCO value for each month of the year 2022. This is given by the sum of the effect of covariates and the spatiotemporal random effects of Model 3.}
\label{pred_mon_22}
\end{figure}

\begin{figure}[H]
\centering
\includegraphics[width=1\linewidth]{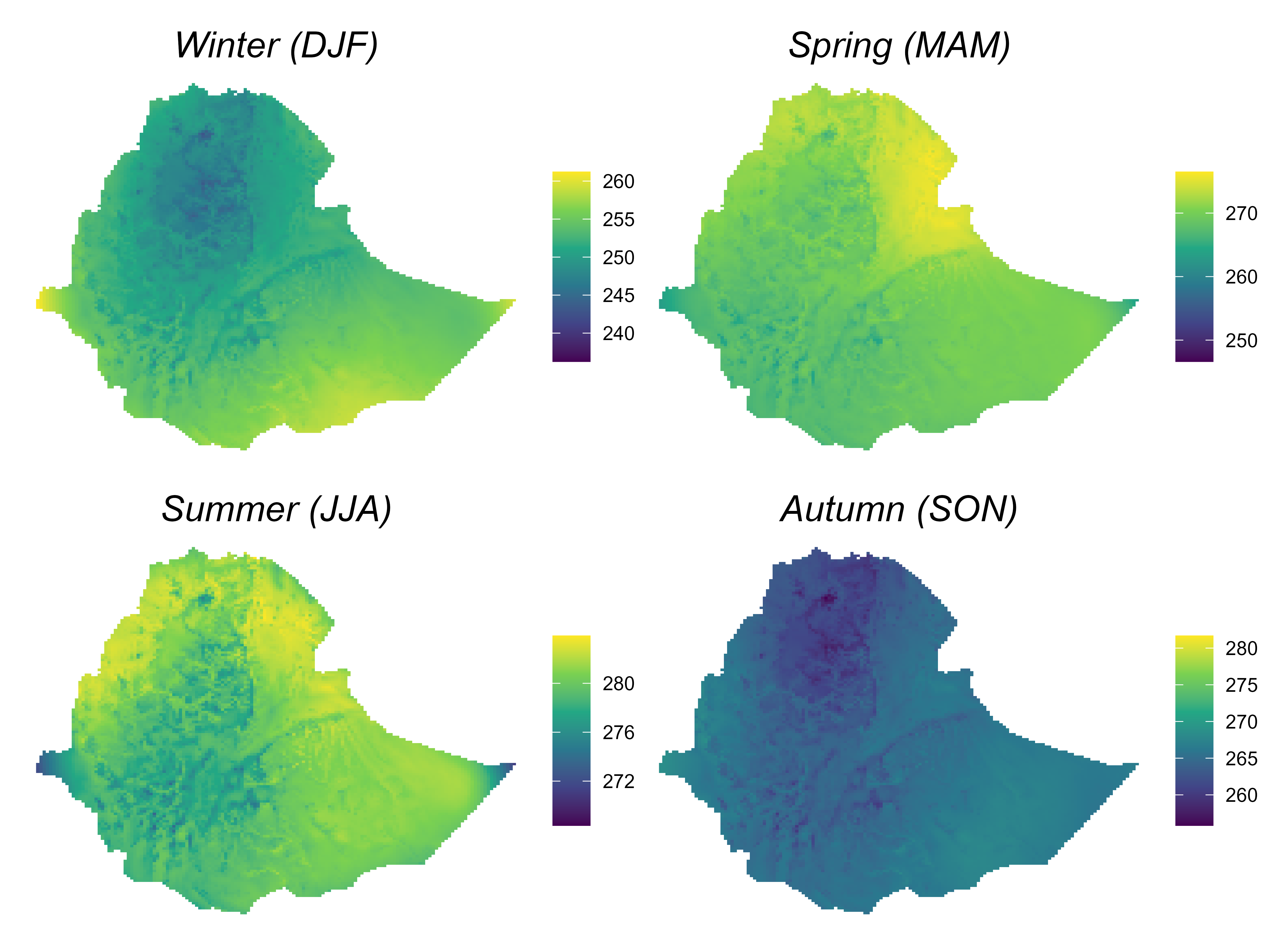} 
\caption{Predictive posterior seasonal mean. The figure shows the Bayesian estimation of the predicted mean of the TCO value for the season of the year 2022. This is given by the sum of the effect of covariates and the spatiotemporal random effects of Model 3.}
\label{pred_seas_22}
\end{figure}

In this section, we focus on the spatial prediction; particularly, we provide monthly maps of TCO variability using 11,967 grids over Ethiopia for the entire year of 2022. Figure \ref{pred_mon_22} shows the monthly predicted posterior mean of TCO concentration in the year 2022. Note that all plots have different color scales. A visual inspection of Figure \ref{pred_mon_22} highlights that relatively higher TCO values have been seen in the months of June to September. Over Ethiopia, TCO typically reaches its seasonal maximum during the June--September (JJAS, kiremt) rainy seasons and its minimum during December--February (DJF) and has moderate values during the fall season of March--May (MAM, belg). This monthly variation is influenced by the stratospheric dynamics, QBO, annual cycle of solar radiation and circulation pattern, and other climatic factors. These results, in turn, depend on the monthly TCO concentration described through the box plots of Figure \ref{boxp} (left). Moreover, the predicted seasonal variation plot is shown in Figure \ref{pred_seas_22}.
%

\section{Discussion and conclusion} \label{conclusion}
This paper proposes a Bayesian spatiotemporal model to investigate total column ozone (TCO) variability over Ethiopia from 2012 to 2022. Unlike prior studies limited to specific sites, our model integrates spatial, temporal, and covariate effects to assess the climatological, meteorological, and topographical drivers of ozone variability across the entire region. The spatiotemporal random effects confirm both spatial and temporal heterogeneity. Notably, clustering patterns are evident in the west, northeast, southeast, and central highlands, likely driven by regional differences in climate and elevation. The model also captures a clear seasonal pattern, with increased TCO during summer (June–August) and decreased values in early months (January–March). These residual fluctuations, after accounting for known covariates, suggest the presence of unobserved atmospheric processes or dynamics not explicitly modeled. 

Our model's key findings reveal a complex interplay between local meteorological, atmospheric, and topographical factors in regulating TCO over Ethiopia. Among the predictors, incoming solar radiation (ISR) emerges as the most influential positive predictor, reinforcing the role of photochemical ozone production in equatorial regions. Stratospheric temperature and the Quasi-Biennial Oscillation (QBO) also show positive associations, consistent with their known roles in thermal stabilization and vertical ozone transport. Conversely, total column water vapor, precipitation, humidity, altitude, and surface temperature exhibit negative associations, implicating these variables in enhanced ozone destruction via water vapor, fueled HO\textsubscript{x} chemistry, intensified stratosphere–troposphere exchange, and radiative cooling. Wind speed lacks statistical significance, suggesting limited influence under regional conditions. Together, these effects illustrate the balance of local photochemical generation and large-scale atmospheric transport, aligning with the equatorial ozone paradox \cite{cohen2014drives}, in which ozone-rich air is transported poleward via the Brewer-Dobson circulation \cite{butchart2014brewer}, limiting local accumulation.

Despite these advances, several limitations remain. The model omits key predictors such as aerosols, dust, ENSO variability \cite{acp-22-15729-2022}, and halogen chemistry, potentially underestimating ozone loss in polluted regions. Assumptions of linearity also preclude capturing nonlinear effects (e.g., QBO–ISR interactions). Future research should integrate high-resolution datasets and employ machine learning techniques to explore nonlinear relationships and interaction terms, as suggested in \cite{alahmadi2024joint}. Extending the framework to jointly model related atmospheric variables using shared-effect approaches \cite{palmi2022bayesian, pan2024spatio} and decomposing trend, seasonal, and cyclical signals with covariates and random effects \cite{laine2014analysing, laurini2019spatio} could yield further insight. 
Moreover, the study of total column ozone over Ethiopia faces broader structural challenges, including limited ground-based monitoring, complex topography, and strong spatiotemporal variability linked to both local meteorology and large-scale climate drivers. These factors introduce uncertainty in both observations and model-based inference. We advocate establishing regional ozone monitoring infrastructure, improving satellite product assimilation, and incorporating stratospheric dynamics and elevation-based variables. Future research should look at the significance of interannual climate trends and work toward region-specific models that can better capture ozone variability across Ethiopia. 
In addition, spatial clustering patterns observed across Ethiopia suggest that localized processes may influence ozone dynamics differently across regions. Incorporating cluster-specific modeling by ozone regimes could help identify subregional differences in covariate effects and better target policy or monitoring interventions. As noted in recent aerosol studies, future research should also investigate the spatiotemporal clustering patterns observed in the northeast, southeast, and central highlands, which may be associated with aerosol loading in the troposphere and stratosphere \cite{ALEMU2024109085}. This points to the importance of integrating aerosol–ozone interactions in regional analyses to better understand the chemical and physical mechanisms driving TCO variability.

Ethiopia’s equatorial latitude and elevated terrain make it a critical zone for both ozone generation and UV exposure. Our model quantifies how atmospheric dynamics and local geography interact to shape TCO variability. While the Montreal Protocol continues to foster global ozone recovery, regional policy must address current vulnerabilities, especially in high-altitude urban centers. Public health measures, UV monitoring, and climate-resilient planning are essential. By aligning atmospheric science with policy and sustainable development, Ethiopia has the opportunity to lead in tackling the interlinked challenges of ozone depletion and UV exposure in tropical regions.
%

%

\section*{Data Availability Statement}
The data used in this study are publicly available. The atmospheric and meteorological variables were obtained from the ERA5 reanalysis dataset provided by the European Centre for Medium-Range Weather Forecasts (ECMWF) through the Copernicus Climate Data Store (\url{https://cds.climate.copernicus.eu}). Total Column Ozone (TCO) data were retrieved from the Ozone Mapping and Profiler Suite (OMPS) aboard the Suomi NPP satellite, available via NASA’s Earthdata portal (\url{https://earthdata.nasa.gov/}). The cleaned datasets and analysis code used in this study are available from the corresponding author upon reasonable request.

\section*{Acknowledgment}
Yassin Tesfaw Abebe acknowledges support from the International Mathematical Union (IMU) and the Graduate Research Assistantships in Developing Countries (GRAID) Program. 
Lassi Roininen was supported by the Research Council of Finland (grant numbers 353095 and 359183).
We gratefully acknowledge the National Aeronautics and Space Administration (NASA) for providing the OMPS data, accessed via the Goddard Earth Sciences Data and Information Services Center (GES DISC). We also acknowledge the Copernicus Climate Change Service and the European Centre for Medium-Range Weather Forecasts (ECMWF) for providing ERA5 data through the Climate Data Store.

\section*{Disclosure statement}
The authors declare no potential conflict of interest.
%

\bibliographystyle{plainnat}
\bibliography{refs_tco2}

\newpage
%


\end{document}